\documentclass{article}

\usepackage{afterpage}
\newcommand{\pointlist}{\textsf{pointList}}
\newcommand{\hasNoFreeValue}{\textsf{hasNoFreeValue}}
\newcommand{\truncate}{\textsf{truncate}}
\newcommand{\depth}{\textsf{depth}}
\newcommand{\isFull}{\textsf{backtracked}}
\newcommand{\getFreeValue}{\textsf{getFreeValue}}
\newcommand{\seekGap}{\textsf{seekGap}}
\newcommand{\setFrontier}{\textsf{setFrontier}}
\newcommand{\frontier}{\textsc{curFrontier}}

\newcommand{\pp}{free tuple\xspace}
\newcommand{\pps}{free tuples\xspace}
\newcommand{\dbox}[1]{\langle #1 \rangle}
\newcommand{\vars}{\textnormal{vars}}
\newcommand{\calD}{\mathcal D}
\newcommand{\Dom}{\mathbb N}

\newcommand{\agm}{\textsf{AGM}\xspace}
\usepackage{url}
\usepackage{rotating}
\usepackage{multirow}
\usepackage{rotating}
\usepackage{cite}
\usepackage{color}
\usepackage{xspace}
\usepackage{balance}
\usepackage{kpfonts}
\usepackage{latexsym}
\usepackage{todonotes}
\usepackage{amsmath}
\usepackage{amsfonts}
\usepackage{graphicx}
\usepackage{tikz}
\usepackage{tikz-3dplot}
\usepackage{subfigure}
\usepackage[T1]{fontenc}
\usepackage{times}
\usepackage[noend]{algpseudocode}
\usepackage[margin=1in]{geometry}
\usepackage[matha]{mathabx}
\usepackage{mathtools}
\usepackage{txfonts} 
\usepackage{hyperref}

\newcommand{\eat}[1]{}

\def\compactify{\itemsep=0pt \topsep=0pt \partopsep=0pt \parsep=0pt}
\let\latexusecounter=\usecounter

\usepackage{algorithm}

\usepackage{amsthm}

\newcommand{\la}{\leftarrow}

\newcommand{\be}{\begin{enumerate}}
\newcommand{\ee}{\end{enumerate}}
\newcommand{\bi}{\begin{itemize}}
\newcommand{\ei}{\end{itemize}}
\newcommand{\beq}{\begin{equation}}
\newcommand{\eeq}{\end{equation}}

\newcommand{\bp}{\begin{proof}}
\newcommand{\ep}{\end{proof}}
\newcommand{\bcor}{\begin{cor}}
\newcommand{\ecor}{\end{cor}}
\newcommand{\bthm}{\begin{thm}}
\newcommand{\ethm}{\end{thm}}
\newcommand{\blmm}{\begin{lmm}}
\newcommand{\elmm}{\end{lmm}}
\newcommand{\bdefn}{\begin{defn}}
\newcommand{\edefn}{\end{defn}}
\newcommand{\bprop}{\begin{prop}}
\newcommand{\eprop}{\end{prop}}
\newcommand{\bconj}{\begin{conj}}
\newcommand{\econj}{\end{conj}}
\newcommand{\bopm}{\begin{opm}}
\newcommand{\eopm}{\end{opm}}
\newcommand{\brmk}{\begin{rmk}}
\newcommand{\ermk}{\end{rmk}}

\newcommand{\suchthat}{\ | \ }

\newcommand{\mv}[1]{\mathbf{#1}}

\newtheorem{thm}{Theorem}[section]
\newtheorem{lmm}[thm]{Lemma}
\newtheorem{prop}[thm]{Proposition}
\newtheorem{cor}[thm]{Corollary}

\theoremstyle{definition}
\newtheorem{idea}{Idea}

\newtheorem{defn}[thm]{Definition}
\newtheorem{opm}[thm]{Open Problem}
\newtheorem{conj}[thm]{Conjecture}

\newtheorem{rmk}[thm]{Remark}

\newlength{\toppush}
\def\compactify{\itemsep=0pt \topsep=0pt \partopsep=0pt \parsep=0pt}
\let\latexusecounter=\usecounter

\def\compactifytwo{\itemsep=-1pt \topsep=-1pt \partopsep=-2pt \parsep=-1pt
\labelwidth=-2pt \leftmargin=-10pt}

\newcommand{\LFTJ}{\textsf{LFTJ}\xspace}

\newcommand{\nxt}{\textsf{Next}}

\newcommand{\ins}{\textsf{insert}}
\newcommand{\insconst}{\textsc{InsConstraint}}

\newcommand{\chld}{\textsf{child}}

\newcommand{\intv}{\textsc{intervals}}
\newcommand{\ctree}{\textsc{ConstraintTree}}

\newcommand{\getpp}{\textsf{computeFreeTuple}}
\newcommand{\getPP}{\textsf{computeFreeTuple}\xspace}

\newcommand{\pattern}{\textsf{pattern}}

\newcommand{\outspace}{\mathcal O}

\newcommand{\atoms}{\mathrm{atoms}}

\newcommand{\ms}{\text{Minesweeper}\xspace}
\newcommand{\lftj}{\text{Leapfrog Triejoin}\xspace}
\newcommand{\cds}{\text{CDS}\xspace}
\newcommand{\cert}{\mathcal C}
\newcommand{\boxcert}{\mathcal C_{\Box}}

\newcommand{\calH}{\mathcal H}

\newcommand{\calV}{\mathcal V}
\newcommand{\calE}{\mathcal E}

\renewenvironment{quote}{%
  \list{}{%
    \leftmargin0.5cm   
    \rightmargin\leftmargin
  }
  \item\relax
}
{\endlist}

\title{
  Join Processing for Graph Patterns: \\
  An Old Dog with New Tricks
}

\author{Dung Nguyen$^1$ \and Molham
  Aref$^1$ \and Martin Bravenboer$^1$ \and George Kollias$^1$ \and Hung Q. Ngo$^2$ \and
  Christopher R\'e$^3$ \and Atri Rudra$^2$ \and \\
  $^1$LogicBlox, $^2$SUNY Buffalo, and $^3$Stanford
}
\begin{document}
\maketitle

\begin{abstract}
Join optimization has been dominated by Selinger-style, pairwise
optimizers for decades. But, Selinger-style algorithms are
asymptotically suboptimal for applications in graphic analytics. This
suboptimality is one of the reasons that many have advocated
supplementing relational engines with specialized graph processing
engines. Recently, new join algorithms have been discovered that
achieve optimal worst-case run times for any join or even so-called
beyond worst-case (or instance optimal) run time guarantees for
specialized classes of joins. These new algorithms match or improve on
those used in specialized graph-processing systems. This paper asks
{\it can these new join algorithms allow relational engines to close
  the performance gap with graph engines?}

We examine this question for graph-pattern queries or join queries. We
find that classical relational databases like Postgres and MonetDB or
newer graph databases/stores like Virtuoso and Neo4j may be orders of
magnitude slower than these new approaches compared to a fully
featured RDBMS, LogicBlox, using these new ideas. Our results
demonstrate that an RDBMS with such new algorithms can perform as well
as specialized engines like GraphLab -- while retaining a high-level
interface. We hope our work adds to the ongoing debate of the role of
graph accelerators, new graph systems, and relational systems in
modern workloads.
\end{abstract}

\section{Introduction}

For the last four decades, Selinger-style pairwise-join enumeration
has been the dominant join optimization
paradigm~\cite{Selinger:1979:APS:582095.582099}.  Selinger-style
optimizers are designed for joins that do not filter tuples, e.g.,
primary-key-foreign-key joins that are common in OLAP. Indeed, the
result of a join in an OLAP query plan is often no smaller than either
input relation. In contrast, in graph applications, queries search for
structural patterns, which filter the data. These regimes are
quite different, and not surprisingly there are separate OLAP and
graph-style systems. Increasingly, analytics workloads contain both
traditional reporting queries and graph-based
queries~\cite{rudolf2013graph}. Thus, it would be desirable to have
one engine that is able to perform well for join processing in both of
these different analytics settings.

Unifying these two approaches is challenging both practically and
theoretically. Practically, graph engines offer orders of magnitude
speedups over traditional relational engines, which has led many to
conclude that these different approaches are irreconcilable. Indeed,
this difference is fundamental: recent theoretical results suggest
that Selinger-style, pair-wise join optimizers are {\em asymptotically
  suboptimal} for this graph-pattern
queries~\cite{NPRR,leapfrog,AGM08}. The suboptimality lies in the fact
that Selinger-style algorithms only consider pairs of joins at a time,
which leads to runtimes for cyclic queries that are asymptotically
slower by factors in the size of the data, e.g.,
$\Omega(\sqrt{N})$-multiplicative factor worse on a database with $N$
tuples. Nevertheless, there is hope of unifying these two approaches
for graph pattern matching as, mathematically, {\it (hyper)graph
  pattern matching is equivalent to join processing}. Recently,
algorithms have been discovered that have strong theoretical
guarantees, such as optimal runtimes in a worst-case sense or even
instance optimally. As database research is about three things:
performance, performance, and performance, the natural question is:
\begin{quote} 
{\em To what extent do these new join algorithms speed up graph
  workloads in an RDBMS?}
\end{quote}

To take a step toward answering this question, we embed these new join
algorithms into the LogicBlox (LB) database engine, which is a
fully featured RDBMS that supports rich queries, transactions, and
analytics. We perform an experimental comparison focusing on a broad
range of analytics workloads with a large number of state-of-the-art
systems including row stores, column stores, and graph processing
engines. Our  technical contribution is the first empirical
evaluation of these new join algorithms. At a high-level, our message
is that these new join algorithms provide substantial speedups over
both Selinger-style systems and graph processing systems for some data
sets and queries. Thus, they may require further investigation for
join and graph processing.

We begin with a brief, high-level description of these new
algorithms, and then we describe our experimental results.

\paragraph*{An Overview of New Join Algorithms}
Our evaluation focuses on two of these new-style algorithms, LeapFrog
TrieJoin (LFTJ)~\cite{leapfrog}, a worst-case optimal algorithm, and
Minesweeper (MS), a recently proposed ``beyond-worst-case''
algorithm~\cite{nnrr}.

LFTJ is a multiway join algorithm that transforms the join into a
series of nested intersections. LFTJ has a running time that is
worst-case optimal, which means for every query there is some family
of instances so that any join algorithm takes as least much time as
LFTJ does to answer that query. These guarantees are non-trivial; in
particular, any Selinger-style optimizer is slower by a factor that
depends on the size of the data, e.g., by a factor of
$\Omega(\sqrt{N})$ on a database with $N$ tuples. Such algorithms were
discovered recently~\cite{NPRR,leapfrog}, but LFTJ has been in
LogicBlox for several years.

Minesweeper's main idea is to keep careful track of every comparison with
the data to infer where to look next for an output.  This allows
Minesweeper to achieve a so-called ``beyond worst-case guarantee''
that is substantially stronger than a worst-case running time
guarantees: for a class called comparison-based algorithms, containing
all standard join algorithms including LFTJ, beyond worst-case
guarantees that {\em any} join algorithm takes no more than a constant
factor more steps on {\em any} instance. Due to indexing, the runtime
of some queries can even be {\em sublinear} in the size of the
data. However, these stronger guarantees only apply for a limited
class of acyclic queries (called $\beta$-acyclic).

\paragraph*{Benchmark Overview} 
The primary contribution of this paper is a benchmark of these new
style algorithms against a range of competitor systems on
graph-pattern matching workloads. To that end, we select a 
traditional row-store system (Postgres), a column-store system (MonetDB)
and graph systems (virtuoso,
neo4j, and graphlab). We find that for cyclic queries these new join
systems are substantially faster than relational systems and
competitive with graph systems. We find that LFTJ is performant in
cyclic queries on dense data, while \ms is superior for acyclic
queries.

\paragraph*{Contributions} This paper makes two contributions:
\begin{enumerate}\compactify
\item We describe the first practical implementation of a beyond
  worst-case join algorithm.

\item We perform the first experimental validation that describes
  scenarios for which these new algorithms are competitive with
  conventional optimizers and graph systems.
\end{enumerate}

\section{Background}

We recall background on graph patterns, join processing, and
hypergraph representation of queries along with the two new join
algorithms that we consider in this paper.

\subsection{Join query and hypergraph representation}

A (natural) {\em join query} $Q$ is
specified by a finite set $\atoms(Q)$ of relational symbols, denoted
by $Q = \ \Join_{R \in \atoms(Q)} R$. Let $\vars(R)$ denote
the set of attributes in relation $R$, and $$\vars(Q) =
\bigcup_{R\in\atoms(Q)} \vars(R).$$
Throughout this paper, let $n = |\vars(Q)|$ and $m = |\atoms(Q)|$.
For example, in the following so-called {\em triangle query}
\[ Q_\triangle = R(A, B) \Join S(B, C) \Join T(A, C). \]
we have $\vars(Q_\triangle) = \{A, B, C\}$, $\vars(R) = \{A, B\}$,
$\vars(S) = \{B, C\}$, and $\vars(T) = \{A, C\}$. The structure of a
join query $Q$ can be represented by a hypergraph $\calH(Q) = (\calV,
\calE)$, or simply $\calH = (\calV, \calE)$.  The vertex set is $\calV
= \vars(Q)$, and the edge set is defined by
\[ \calE = \left\{ \vars(R) \suchthat R \in \atoms(Q) \right\}. \]
Notice that if the query hypergraph is exactly finding a pattern in a
graph. For so-called {\em
  $\alpha$-acyclicity}~\cite{DBLP:journals/jacm/Fagin83,DBLP:books/aw/AbiteboulHV95},
the celebrated Yannakakis algorithm~\cite{DBLP:conf/vldb/Yannakakis81}
runs in linear-time (in data complexity).  On graph databases with
binary relations, both $\alpha-$ and $\beta-$ acyclic can simply be
thought of as the standard notion of acyclic.

\paragraph*{Worst-case Optimal Algorithm}
Given the input relation sizes, Atserias, Grohe, and Marx \cite{AGM08}
derived a linear program that could be used to upper bound the
worst-case (largest) output size in number of tuples of a join
query. For a join query, $Q$ we denote this bound $\agm(Q)$. For
completeness, we describe this bound in
Appendix~\ref{app:agm}. Moreover, this bound is tight in the sense
that there exists a family of input instances for any $Q$ whose output
size is $\Omega(\agm(Q))$. Then, Ngo, Porat, R\'e, and
Rudra (NPRR)~\cite{NPRR} presented an algorithm whose runtime matches the
bound. This algorithm is thus {\em worst-case optimal}. Soon after,
Veldhuizen \cite{leapfrog} used a similar analysis to show that
the \LFTJ algorithm -- a simpler algorithm already implemented in
LogicBlox Database engine -- is also worst-case optimal. A simpler
exposition of these algorithms was described~\cite{skew} and formed
the basis of a recent system~\cite{2015arXiv150302368A}.

\paragraph*{Beyond Worst-case Results}
Although NPRR may be optimal for worst-case instances, there are
instances on which one can improve its runtime. To that end, the
tightest theoretical guarantee is so-called {\em instance optimality}
which says that the algorithm is up to constant factors no slower than
{\em any algorithm}, typically, with respect to some class of
algorithms. These strong guarantees had only been known for restricted
problems~\cite{DBLP:conf/pods/FaginLN01}. Recently, it was shown that
if the query is {\em $\beta$-acyclic}, then a new algorithm called \ms
(described below) is $\log$-instance optimal\footnote{Instance optimal
  up to a logarithmic factor in the database size~\cite{nnrr}. This
  factor is unavoidable.} with respect to the class of
comparison-based joins, a class which includes essentially all known
join and graph processing algorithms. Theoretically, this result is
much stronger--but this algorithm's performance not been previously
reported in the literature.

\subsection{The \lftj Algorithm}

We describe the \lftj algorithm. The main idea is to ``leapfrog'' over
large swaths of tuples that cannot produce output. To describe it, we
need some notation. For any relation $R$, an attribute $A\in
\vars(R)$, and a value $a \in \pi_A(R)$, define
\[ R[a] = \pi_{\vars(R)-\{A\}}(\sigma_{A=a}(R)). \]
That is $R[a]$ is the set of all tuples from $R$ whose $A$-value is
$a$.

At a high-level, the \lftj algorithm can be presented recursively as
shown in Algorithm~\ref{algo:lftj}. In the actual implementation, we
implement the algorithm using a simple iterator interface, which
iterates through tuples. Please see Section~\ref{sec:LB} for more detail.

\begin{algorithm}
\caption{High-level view: \lftj ($\LFTJ(Q)$)}
\label{algo:lftj}
\begin{algorithmic}[1]
\Require{All relations $R$ are in $\atoms(Q)$}
\State Let $\vars(Q) = (A_1,\dots,A_n)$
\State $L \la \bigcap_{R: A_1 \in \vars(R)} \pi_{A_1}(R)$
\If {$n=1$}
\State \Return $L$
\EndIf
\For {each $a_1 \in L$}
  \State \Comment{The following forms a new query on $(A_2,\dots,A_n)$}
  \State $Q_{a_1} \la \bigl(\Join_{R: A_1\in\vars(R)} R[a_1]
  \Join \Join_{R: A_1 \notin \vars(R)} R\bigr)$ \\
  \State $S_{a_1} = \LFTJ(Q_{a_1})$
  \Comment{Recursive call}
\EndFor
\State \Return $\bigcup_{a_1\in L} \{a_1\} \times S_{a_1}$
\end{algorithmic}
\end{algorithm}

One can show that the \lftj (\LFTJ) algorithm runs in time $\tilde O(N
+ \agm(Q))$ for any query $Q$~\cite{leapfrog}. The simplified version
shown above appeared later~\cite{skew}.

\subsection{The Minesweeper Algorithm}

Consider the set of tuples that could be returned by a join (i.e., the
cross products of all domains). Often many fewer tuples are part of
the output than are not. \ms's exploits this idea to focus on quickly
ruling out where tuples are not located rather than where they
are. \ms starts off by obtaining an arbitrary tuple $\mv t \in
\Dom^{\vars(Q)}$ from the output space, called a {\em \pp} (also
called a {\em probe point} in \cite{nnrr}).  By probing into the
indices storing the input relations, we either confirm that $\mv t$ is
an output tuple or we get back $\tilde O(1)$ ``gaps'' or
multi-dimensional rectangles inside which we know for sure {\em no}
output tuple can reside. We call these rectangles {\em gap boxes}.
The gap boxes are then inserted into a data structure called the {\em
  constraint data structure} (\cds).  If $\mv t$ is an output tuple,
then a corresponding (unit) gap box is also inserted in to the \cds.
The \cds helps compute the next \pp, which is a point in the output
space not belonging to any stored gap boxes.  The \cds is a
specialized cache that ensures that we maximally use the information
we gather about which tuples must be ruled out. The algorithm proceeds
until the entire output space is covered by gap boxes.
Algorithm~\ref{algo:ms-overview} gives a high-level overview of how
\ms works.

\begin{algorithm}
\caption{High-level view: \ms algorithm}
\label{algo:ms-overview}
\begin{algorithmic}[1]
\State $\cds \gets \emptyset$ \Comment{No gap discovered yet}
\While {$\cds$ can find $\mv t$ not in any stored gap}
  \If {$\pi_{\vars(R)}(\mv t) \in R$ for every $R\in \atoms(Q)$}
    \State Report $\mv t$ and insert $\mv t$ as a gap into \cds
  \Else
    \State Query all $R \in \atoms(Q)$ for gaps around $\mv t$
    \State Insert those gaps into \cds
  \EndIf
\EndWhile
\end{algorithmic}
\end{algorithm}

A key idea from~\cite{nnrr} was the proof that the total number of gap
boxes that \ms discovers using the above outline is $\tilde
O(|\cert|)$, where $\cert$ is the minimum set of comparisons that {\em
  any} comparison-based algorithm must perform in order to work
correctly on this join. Essentially all existing join algorithms such as
Block-Nested loop join, Hash-Join, Grace, Sort-merge, index-nested,
PRISM, double pipelined, are comparison-based (up to a $\log$-factor
for hash-join). For $\beta$-acyclic queries, \ms is instance optimal
up to an (unavoidable) $\log N$ factor.
  
\section{LogicBlox Database System}
\label{sec:LB}

The LogicBlox database is a commercial database system that from the
ground up is designed to serve as a general-purpose database system
for enterprise applications. The LogicBlox database is currently
primarily used by partners of LogicBlox to develop applications that
have a complex workload that cannot easily be categorized as either
analytical, transactional, graph-oriented, or
document-oriented. Frequently, the applications also have a
self-service aspect, where an end-user with some modeling expertise
can modify or extend the schema dynamically to perform analyses that
were originally not included in the application.

The goal of developing a general-purpose database system is a
deviation from most current database system development, where the
emphasis is on designing specialized systems that vastly outperform
conventional database systems, or to extend one particular
specialization (e.g. analytical) with reasonable support for a
different specialized purpose (e.g. transactional).

The challenging goal of implementing a competitive general-purpose
database, requires different approaches in several components of a
database system. Join algorithms are a particularly important part,
because applications that use LogicBlox have schemas that resemble
graph as well as OLAP-style schemas, and at the same time have a
challenging transactional load. To the best of our knowledge, no
existing database system with conventional join algorithms can
efficiently evaluate queries over such schemas, and that is why
LogicBlox is using a join algorithm with strong optimality guarantees:
LFTJ.

Concretely, the motivation for implementing new optimal join
algorithms are:
\begin{itemize}
\item No previously existing join algorithm efficiently supports the
  graph queries required in applications. On the other hand,
  graph-oriented systems cannot handle OLAP aspects of applications.

\item To make online schema changes easy and efficient, LogicBlox
  applications use unusually high normalization levels, typically
  6NF. The normalized schemas prevent the need to do surgery on
  existing data when changing the schema, and also helps with
  efficiency of analytical workloads (compare to column stores). A
  drawback of this approach is that queries involve a much larger
  number of tables. Selection conditions in queries typically apply to
  multiple tables, and simultaneously considering all the conditions
  that narrow down the result becomes important.

\item As opposed to the approach of highly tuned in-memory databases
  that fully evaluate all queries on-the-fly~\cite{rudolf2013graph},
  LogicBlox encourages the use of materialized views that are
  incrementally
  maintained~\cite{DBLP:journals/corr/abs-1303-5313}. The incremental
  maintenance of the views under all update scenarios is a challenging
  task for conventional joins, in particular combined with a
  transactional load highly efficient maintenance is
  required~\cite{DBLP:journals/corr/Veldhuizen14}.
\end{itemize}

The LogicBlox database system is designed to be highly modular as a
software engineering discipline, but also to encourage
experimentation. Various components can easily be replaced with
different implementations. This enabled the implementation of
Minesweeper, which we compare to the LFTJ implementation and other
systems in this paper.

\section{\ms implementation}
\label{sec:minesweeper}

\subsection{Global attribute order (GAO)}

\begin{figure}                                                                  
\centerline{\includegraphics[width=3in]{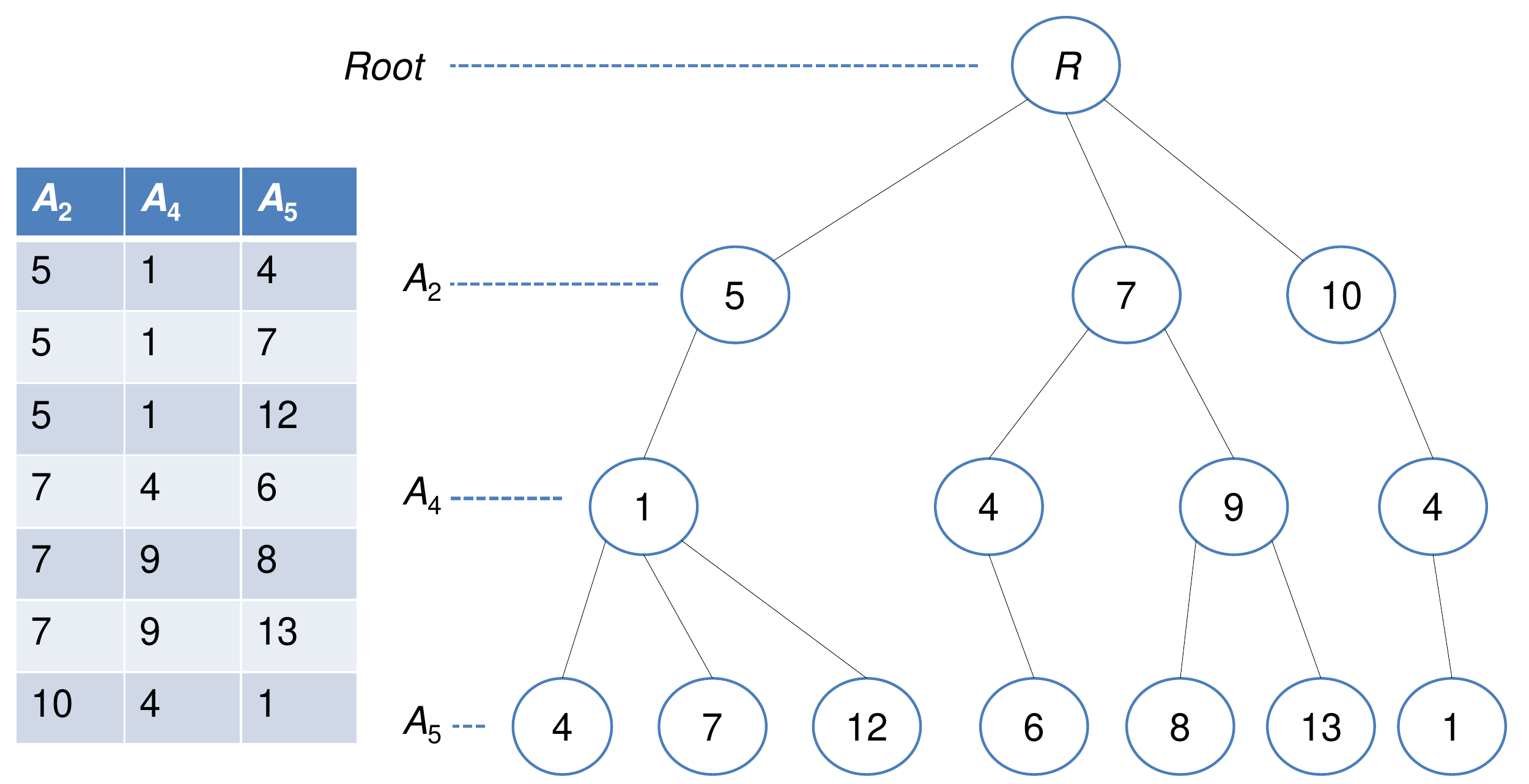}}                           
\caption{How \ms ``views'' input indices}
\label{fig:R-trie}                                                              
\end{figure}

Both \lftj and \ms work on input relations that are already indexed
using a search-tree data structure such as a traditional B-tree 
which is widely used in commercial relational database 
systems~\cite[Ch.10]{Ramakrishnan:2002:DMS:560733}.
For example, Figure~\ref{fig:R-trie} shows the index for a relation
$R$ on attribute set $\vars(R) = \{A_2, A_4, A_5\}$.
This index for $R$ is in the order $A_2,A_4,A_5$.
Furthermore, there is an ordering of all the attributes 
in $\vars(Q)$ -- called the {\em global attribute order} (GAO) --
such that all input relations are indexed consistent with this GAO.
This assumption shall be referred to as the {\em GAO-consistency assumption}.
For example, for the triangle query $Q = R(A,B) \Join S(B, C) \Join T(A, C)$,
if the GAO is $B, A, C$, then
$R$ is indexed in the $(B, A)$ order,
$S$ in the $(B, C)$, and $T$ in the $(A, C)$.

\subsection{Gap boxes, constraints, and patterns}

To describe the gap boxes, let us consider an example.
Suppose $\vars(Q) = \{A_0, A_1,\dots,A_6\}$ with 
$(A_0, A_1,A_2,\dots,A_6)$ being the GAO. Suppose the relation $R$ shown in 
Figure~\ref{fig:R-trie}
is an input relation. Consider the following \pp
\[ \mv t = (t_0, t_1,t_2,t_3,t_4,t_5,t_6) = (2, 6, 6, 1, 3, 7, 9). \]
We first project this tuple down to the coordinate subspace spanned by the
attributes of $R$: $(t_2, t_4, t_5) = (6, 3, 7)$.
From the index structure for $R$, we see that $t_2 = 6$ falls between
the two $A_2$-values $5$ and $7$ in the relation. Thus, this index returns a
gap consisting of all points lying between the two hyperplanes $A_2=5$ and
$A_2=7$. 
This gap is encoded with the constraint 
\begin{equation}\label{eqn: c}
\mv c = \langle *, *, (5, 7), *, *, *, * \rangle,
\end{equation}
where $*$ is the wildcard character matching any value in the corresponding
domain, and $(5,7)$ is an open interval on the $A_2$-axis.
On the other hand, suppose the \pp is
\[ \mv t = (t_0, t_1,t_2,t_3,t_4,t_5,t_6) = (2, 6, 7, 1, 5, 8, 9). \]
Then, a gap returned might be the band in the hyperplane $A_2=7$,\;
$4 < A_4 < 9$. The encoding of this gap is
\begin{equation}\label{eqn: c'}
\mv c' = \langle *, *, 7, *, (4, 9), *, * \rangle.
\end{equation}
The number $7$ indicates that this gap is inside the hyperplane $A_2=7$,
and the open interval encodes all points inside this hyperplane where
$4 < A_4 < 9$.

Due to the GAO-consistency assumption, all the constraints returned by the input
indices have the property that for each constraint there is only one interval 
component, after that there are only wildcard component. 

\begin{defn}[Constraint and pattern]
Gap boxes are encoded by {\em constraints}. 
Each constraint is an $n$-dimensional tuple $\mv c = \dbox{c_0,\dots,c_{n-1}}$,
where each $c_i$ is either a member of $\Dom \cup \{*\}$ or an {\em open
interval} $(\ell, r)$ where $\ell, r \in \Dom$. For each constraint $\mv c =
\dbox{c_0,\dots,c_{n-1}}$, there is only one $c_i$
which is an interval, after which all components are wildcards.
The tuple of components {\em before} the interval component is called
the {\em pattern} of the constraint, denoted by $\pattern(\mv c)$. 
For example, $\pattern(\mv c) = \dbox{*, *}$ for the constraint $\mv c$ defined 
in \eqref{eqn: c}, and $\pattern(\mv c') = \dbox{*, *, 7, *}$ for the constraint
in \eqref{eqn: c'}.
\end{defn}

\subsection{The constraint data structure (\cds)}
\label{sec:cds-outline}

\begin{figure}[ht]

\begin{center}
\begin{tikzpicture}


\draw (0,0) circle (.25);
\draw (0,-1) circle (.25);
\draw (0,-2) circle (.25);

\draw (0,-.25) -- (0,-.75);
\draw (0,-1.25) -- (0,-1.75);

\node [right] at (0,-.5) {{\scriptsize $*$}};
\node [right] at (0,-1.5) {{\scriptsize $*$}};

\node [right] at (.25,-2) {{\scriptsize $(5,7)$}};

\draw [gray, dotted] (-1,0) -- (1,0);
\node [left] at (-1,0) {\textcolor{gray}{\scriptsize $A_0$}};
\draw [gray, dotted] (-1,-1) -- (1,-1);
\node [left] at (-1,-1) {\textcolor{gray}{\scriptsize $A_1$}};
\draw [gray, dotted] (-1,-2) -- (1,-2);
\node [left] at (-1,-2) {\textcolor{gray}{\scriptsize $A_2$}};


\begin{scope}[shift={(4,0)}]
\draw (0,0) circle (.25);
\draw (0,-1) circle (.25);
\draw (0,-2) circle (.25);
\draw (0,-3) circle (.25);
\draw (0,-4) circle (.25);

\draw (0,-.25) -- (0,-.75);
\draw (0,-1.25) -- (0,-1.75);
\draw (0,-2.25) -- (0,-2.75);
\draw (0,-3.25) -- (0,-3.75);

\node [right] at (0,-.5) {{\scriptsize $*$}};
\node [right] at (0,-1.5) {{\scriptsize $*$}};
\node [right] at (0,-2.5) {{\scriptsize $7$}};
\node [right] at (0,-3.5) {{\scriptsize $*$}};

\node [right] at (.25,-2) {{\scriptsize $(5,7)$}};
\node [right] at (.25,-4) {{\scriptsize $(4,9)$}};

\draw [gray, dotted] (-1,0) -- (1,0);
\node [left] at (-1,0) {\textcolor{gray}{\scriptsize $A_0$}};
\draw [gray, dotted] (-1,-1) -- (1,-1);
\node [left] at (-1,-1) {\textcolor{gray}{\scriptsize $A_1$}};
\draw [gray, dotted] (-1,-2) -- (1,-2);
\node [left] at (-1,-2) {\textcolor{gray}{\scriptsize $A_2$}};
\draw [gray, dotted] (-1,-3) -- (1,-3);
\node [left] at (-1,-3) {\textcolor{gray}{\scriptsize $A_3$}};
\draw [gray, dotted] (-1,-4) -- (1,-4);
\node [left] at (-1,-4) {\textcolor{gray}{\scriptsize $A_4$}};

\end{scope}


\begin{scope}[shift={(2,-5)}]


\draw (0,0) circle (.25);
\draw (0,-1) circle (.25);
\draw (0,-3) circle (.25);
\draw (1,-2) circle (.25);
\draw (-1,-2) circle (.25);
\draw (2,-3) circle (.25);
\draw (-2,-3) circle (.25);
\draw (-1,-4) circle (.25);
\draw (1,-4) circle (.25);
\draw (2,-4) circle (.25);


\draw [blue] (0,-.25) -- (0,-.75);
\draw [blue] (0,-1.25) -- (-1,-1.75);
\draw (0,-1.25) -- (1,-1.75);
\draw (-1,-2.25) -- (-2,-2.75);
\draw [blue] (-1,-2.25) -- (0,-2.75);
\draw (1,-2.25) -- (2,-2.75);
\draw [blue] (0,-3.25) -- (-1,-3.75);
\draw (0,-3.25) -- (1,-3.75);
\draw (2,-3.25) -- (2,-3.75);


\node [right] at (0,-.5) {\textcolor{blue}{\scriptsize $*$}};
\node [above] at (-.5,-1.5) {\textcolor{blue}{\scriptsize $1$}};
\node [above] at (.5,-1.5) {{\scriptsize $*$}};
\node [above] at (-1.5,-2.5) {{\scriptsize $2$}};
\node [above] at (-.5,-2.5) {\textcolor{blue}{\scriptsize $3$}};
\node [above] at (-.5,-3.5) {\textcolor{blue}{\scriptsize $5$}};
\node [above] at (.5,-3.5) {{\scriptsize $*$}};
\node [below] at (1.5,-2.5) {{\scriptsize $7$}};
\node [right] at (2,-3.5) {{\scriptsize $*$}};


\node at (0,-3) {\textcolor{red}{\scriptsize $u$}};
\node at (-1,-4) {\textcolor{blue}{{\scriptsize $v$}}};
\node at (1,-4) {{\scriptsize $w$}};


\node [left] at (-1.25,-3.5) {\textcolor{blue}{\scriptsize $\pattern(v)=\langle *,1,3,5\rangle$}};


\node [right] at (1.25, -2) {{\scriptsize $(5,7)$}};
\node [right] at (2.25, -4) {{\scriptsize $(4,9)$}};
\node [left] at (.75, -4) {{\scriptsize $(5,10)$}};
\node [left] at (-1.25, -4) {{\scriptsize $(1,3), (3,9), (10,14)$}};
\node [left] at (-2.25,-3) {{\scriptsize $(10,19)$}};
\node [left] at (-1.25,-2) {{\scriptsize $(1,3),(9,10)$}};

\draw [blue, dotted] (-1.3, -4.2) rectangle (-3.5,-3.8);
\node [below] at (-2.4,-4.2) {\textcolor{blue}{\scriptsize $v.\intv$}};

\node at (0,-5) {{\scriptsize $(1,L), (3,L\&R), (9,R), (10,L), (14,R)$}};
\draw [blue, dotted] (-2,-5.2) rectangle (2,-4.8);
\node [below] at (0,-5.2) {\textcolor{blue}{\scriptsize $v.\pointlist$}};
\draw [blue, dotted, ->] (-2.4,-4.4) -- (-2.4,-5) -- (-2,-5);


\node [right] at (-3.9,-.3) {\textcolor{red}{\scriptsize $u.\chld[5]=v$}};
\node [right] at (-3.9,-.6) {\textcolor{red}{\scriptsize $u.\chld[*]=w$}};


\draw [gray, dotted] (-4,0) -- (3,0);
\node [left] at (-4,0) {\textcolor{gray}{\scriptsize $A_0$}};
\draw [gray, dotted] (-4,-1) -- (3,-1);
\node [left] at (-4,-1) {\textcolor{gray}{\scriptsize $A_1$}};
\draw [gray, dotted] (-4,-2) -- (3,-2);
\node [left] at (-4,-2) {\textcolor{gray}{\scriptsize $A_2$}};
\draw [gray, dotted] (-4,-3) -- (3,-3);
\node [left] at (-4,-3) {\textcolor{gray}{\scriptsize $A_3$}};
\draw [gray, dotted] (-4,-4) -- (3,-4);
\node [left] at (-4,-4) {\textcolor{gray}{\scriptsize $A_4$}};

\end{scope}
\end{tikzpicture}
\end{center}

\vskip-1em
\caption{Example of the $\ctree$ data structure for $n=5$. The top left figure 
is the CDS after the constraint $\langle *,*,(5,7),*,*\rangle$ is inserted into 
an empty CDS. The top right is the CDS after $\langle *,*,7,*,(4,9)\rangle$ was 
added. The bottom figure is the CDS after the 
following constraints are further added:
$\langle *,1,(1,3),*,*\rangle,
\langle *,1,(9,10),*,*\rangle,
\langle *,1,2,(10,19),*\rangle,
\langle *,1,3,5,(3,9)\rangle$,
$\langle *,1,3,5,(1,3)\rangle,
\langle *,1,3,5,(10,14)\rangle,
\langle *,1,3,*,(5,10)\rangle$.}
\label{fig:ctree}

\end{figure}
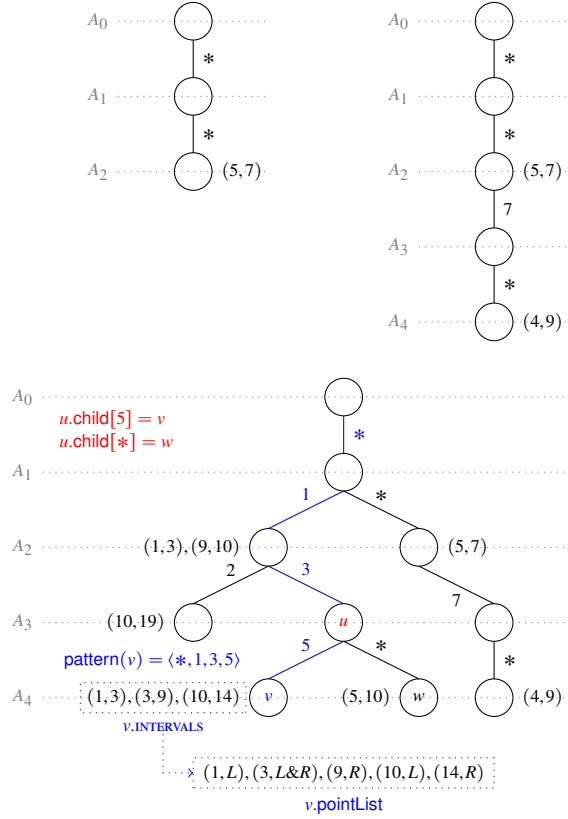

The \cds is a data structure that implements two functions as
efficiently as possible: (1) $\insconst(\mv c)$ takes a new constraint
$\mv c$ and inserts it into the data structure, and (2) $\getpp$
computes a tuple $\mv t$ that does not belong to any constraints (i.e. gap
boxes) that have been inserted into the \cds. $\getpp$ returns {\sc true}
if $\mv t$ was found, and {\sc false} otherwise.

To support these operations, we implement the \cds using a tree
data structure with at most $n$ levels,
one for each of the attributes following the GAO.
Figure~\ref{fig:ctree} illustrates such a tree.

For each node $v \in \cds$, there are two important
associated lists: $v.\chld$ and $v.\intv$.
\begin{itemize}
    \item $v.\chld$ is a map from $\Dom\cup\{*\}$ to children nodes of $v$ in 
        the \cds, where each parent-child edge is labeled with a label in
        $\Dom\cup\{*\}$. In particular, $v.\chld[a]$ is the child node 
        $u$ of $v$, where the $vu$ edge is labeled with $a$.
        Consequently, each node $v$ in the \cds is identified by the tuple of
        labels of edges from the root to $v$. Naturally we call this tuple of
        labels $\pattern(v)$.
        It is certainly possible for $v.\chld = \emptyset$, in which case $v$ is
        a leaf node of the \cds. (See Figure~\ref{fig:ctree} for an
        illustration.)
    \item $v.\intv$ is a set of {\em disjoint} open intervals of the form
        $(\ell, r)$, where $\ell, r \in \Dom \cup \{-\infty, +\infty\}$.
        Each interval $(\ell, r) \in v.\intv$ corresponds to a constraint
        $\mv c = \dbox{\pattern(v), (\ell, r), *, \dots, *}$.
        In particular, when we insert new intervals into $v.\intv$, overlapped
        intervals are automatically merged.
\end{itemize}

\begin{idea}[Point List]
    To implement the above lists, we used one single list for each
    node called the $\pointlist$. For each node $v$ of the \cds, $v.\pointlist$ is
    a (sorted) subset of $\Dom \cup \{-\infty, +\infty\}$. Each value $x \in
    v.\pointlist$ has a set of associated data members, which tells us whether 
    $x$ is a left end point, right end point of some interval in $v.\intv$, or
    both a left end point of some interval and the right end point of another.
    For example, if $(1, 10)$ and $(10, 20)$ are both in $v.\intv$, then $x=10$
    is both a left and a right end point. The second piece of information is a
    pointer to another node of the \cds. If $v.\chld[x]$ exists then this
    pointer points to that child, otherwise the pointer is set to $\null$.
    The $\pointlist$ is made possible by the fact that when newly inserted
    intervals are not only merged with overlapping intervals, but also eliminate
    children nodes whose labels are inside the newly inserted interval.
    We will see later that the $\pointlist$ adds at least two other benefits
    for speeding up \ms and \#\ms.
\end{idea}

With the above structure, inserting a constraint into the \cds is
straightforward. In the next section we describe the \getPP algorithm which is
the most important algorithm to make \ms efficient.
There are two additional simple functions associated with each node $v$ in the \cds
that are used often in implementing \getPP:
\begin{itemize}
    \item $v.\nxt(\texttt{int } x)$ returns the smallest integer $y \geq x$ 
        such that $y$ does {\em not} belong to any interval in the list $v.\intv$.
    \item $v.\hasNoFreeValue()$ returns whether $v.\nxt(-1) = +\infty$, i.e. all values
        in $\Dom$ are covered by intervals in $v.\intv$.
\end{itemize}

\subsection{A slight change to \getPP}

\begin{idea}[The moving frontier]
While basic \ms (Algorithm~\ref{algo:ms-overview}) states that a
new \pp $\mv t$ is computed afresh at each outer loop of the
algorithm, this strategy is inefficient. In our implementation, we move 
\pps in lexicographically increasing order which speeds up the algorithm
because some constraints can be implicit instead of explicitly inserted into the
\cds.

More concretely, the \cds maintains a tuple $\frontier \in \Dom^{n}$
called the {\em frontier}. We start with 
$\frontier = (-1, -1, \dots, -1)$.
In the \getPP function, the \cds has to compute the next
\pp $\mv t$ which is {\em greater than or equal to} the frontier
in lex order {\em and} -- of course -- does not satisfy any of the constraint
stored in the \cds. In particular, all tuples below the frontier are either
implicitly ruled out or are output tuples which have already been reported.
\end{idea}

This idea has a very important benefit. When the current
\pp is verified to be an output tuple, then we do not need to insert a
new constraint just to rule out a single tuple in the output space! 
(If we were to insert such a constraint, it would be of the form
$\dbox{t_0,\dots,t_{n-2},(t_{n-1}-1,t_{n-1}+1)}$, which adds a very significant
overhead in terms of space and time as many new pointers are allocated.)
Instead, we change the frontier to be 
$\frontier = (t_0,\dots,t_{n-2},t_{n-1}+1)$ and ask the \cds for the next
\pp.

\subsection{Obtaining gaps from relations}

\begin{idea}[Geometric certificate]
The key notion of {\em comparison certificate} from \cite{nnrr} was the set $\cert$
of comparisons that any comparison-based algorithm must discover to certify 
the correctness of its output. In order to show that \ms (as outlined in
Algorithm~\ref{algo:ms-overview}) inserts into the \cds $\tilde O(|\cert|)$ many
constraints, the gaps were crafted carefully in \cite{nnrr} so that in each
iteration at least one comparison in $\cert$ is ``caught.''

Let us forget about the Boolean notion of comparison certificate for now
and examine what the \cds sees and processes. The \cds has a set of output 
boxes, and a collection of gap boxes which do not contain any output point. 
When the \cds cannot find a \pp anymore, the union of output boxes 
and gap boxes is the entire output space. In other words, every point in the 
output space is either an output point, or is covered by a gap box. We will 
call the collection of gap boxes satisfying this property a {\em box
certificate}, denoted by $\boxcert$. 
A box certificate is a purely geometric notion, and on the surface does not 
seem to have anything to do with comparisons. Yet
from the results in \cite{nnrr}, we now know that a box certificate of minimum 
size is a lowerbound on the number of comparisons issued by any 
comparison-based join algorithm (with the GAO-assumption).
\end{idea}

With this observation in mind, for every \pp $\mv t$, we only 
need to find from each relation $R$ the {\em maximal} gap box containing the 
projection $\pi_{\vars(R)}(\mv t)$, and if $\pi_{\vars(R)}(\mv t) \in R$ then
no gap box is reported. Suppose $R$ has $k$ attributes
$A_{i_1},\dots,A_{i_k}$, for $i_1<\dots<i_k$, and 
suppose $\pi_{\vars(R)}(\mv t) \notin R$, 
the maximal gap box from $R$ with respect to the \pp $\mv t$
is found as follows. Let $j \in [k]$ be the smallest integer such that
\[ \pi_{A_{i_1},\dots,A_{i_{j-1}}}(\mv t) \in 
    \pi_{A_{i_1},\dots,A_{i_{j-1}}}(R) \text{ and }
   \pi_{A_{i_1},\dots,A_{i_{j}}}(\mv t) \notin 
    \pi_{A_{i_1},\dots,A_{i_{j}}}(R).
\]
To shorten notations, write $J = \pi_{A_{i_1},\dots,A_{i_{j}}}(R)$
and define
\begin{eqnarray*}
    \ell &=& \max \{ x \in \Dom \cup \{-\infty\}
                     \suchthat x < t_{i_j} \wedge
    (t_{i_1},\dots,t_{i_{j-1}},x) \in J \}\\
    r &=& \min \{ x \in \Dom \cup \{+\infty\}
                     \suchthat x > t_{i_j} \wedge
    (t_{i_1},\dots,t_{i_{j-1}},x) \in J \}.
\end{eqnarray*}
Then, the constraint from $R$ is $\mv c_R = \dbox{c_0,\dots,c_{n-1}}$ where
\[
    c_i =
    \begin{cases}
        * & \text{ if } i \notin \{i_1,\dots,i_j\}\\
        t_{i_p} & \text{ if } i = i_p \wedge p < j\\
        (\ell,r) & \text{ if } i = i_j.
    \end{cases}
\]

\begin{idea}[Avoid repeated \seekGap()]\label{idea:seek gap}
To find a constraint as described above from a relation $R$, we use the 
operators \texttt{seek\_lub()} (least upper bound) and \texttt{seek\_glb()}
(greatest lower bound) from LogicBlox' Trie index interface. These operations
are generally costly in terms of runtime as they generally require disk I/O.
Hence, we try to reduce the number of calls to \texttt{seek\_gap()} whenever
possible. In particular, for each constraint inserted into the \cds, we record 
which relation(s) it came from. For example, suppose we have a relation
$R(B,C)$ which has inserted a constraint $\dbox{*, b, (\ell, r)}$ to the
\cds. Then, when the \pp is $\mv t = (a, b, r)$, we knew that no gap
can be found from $R$ without calling any \texttt{seek\_gap()} on $R$.
This simple idea turns out to significantly reduce the overall runtime of
\ms. The speedup when this idea is incorporated is shown in
Table~\ref{table:seek gap}.
\end{idea}

\begin{table*}[!th]
\centerline{
\renewcommand\tabcolsep{5pt}
\def\arraystretch{1.2}
\begin{tabular}{||l||l||r|r|r|r|r|r|r|r|r|r|r|r||}
    \cline{3-14}
    \cline{3-14}
\multicolumn{2}{c|}{-}&\begin{sideways}{ca-GrQc}\end{sideways} &\begin{sideways}p2p-Gnutella04\end{sideways} &\begin{sideways}facebook\end{sideways} &\begin{sideways}ca-CondMat\end{sideways} &\begin{sideways}wiki-vote\end{sideways} &\begin{sideways}p2p-Gnutella31\end{sideways} &\begin{sideways}email-Enron\end{sideways} &\begin{sideways}loc-brightkite\end{sideways} &\begin{sideways}soc-Epinions1\end{sideways} &\begin{sideways}soc-Slashdot0811\end{sideways} &\begin{sideways}soc-Slashdot0902\end{sideways} &\begin{sideways}twitter-combined\end{sideways}\\
\hline
\hline
\multirow{3}{*}{Idea 4}&2-comb&1.38&1.34&1.82&1.54&2.24&1.26&2.06&2.02&2.13&2.30&2.26&2.55\\
&3-path&1.11&1.26&1.72&1.46&1.94&1.24&1.83&1.59&1.93&2.26&2.08&2.54\\
&4-path&1.37&1.45&1.81&1.86&2.06&1.35&2.12&2.10&2.09&2.21&2.19&2.71\\
\hline
\hline
\multirow{3}{*}{Ideas 4\&6}&2-comb&1.51&1.48&2.60&1.69&3.46&1.35&3.30&2.98&3.36&4.16&4.11&4.49\\
&3-path&1.10&1.42&2.13&1.50&2.82&1.27&2.50&1.91&3.13&3.79&3.64&4.52\\
&4-path&1.74&1.49&2.56&2.15&3.34&1.43&3.58&3.26&3.08&3.97&3.93&5.18\\
\hline
\hline
\end{tabular}
}
\caption{Speedup ratio when Ideas~\ref{idea:seek gap} and
  \ref{idea:complete node} are incorporated}
\label{table:seek gap}
\end{table*}

\subsection{\ms' outer loop}

Now that it is clear what the interface to \getPP is and what constraint we
can expect from the input relations, we can make the outer loop of \ms more
precise in Algorithm~\ref{algo:ms-outer}.

\begin{algorithm}[!htp]
\caption{\ms' outer algorithm}
\label{algo:ms-outer}
\begin{algorithmic}[1]
\State $\cds \gets \emptyset$ \Comment{No gap discovered yet}
\State $\cds.\setFrontier(-1,\dots,-1)$ \Comment{Initialize the frontier}
\While {$\cds.\getpp()$} 
  \State $\mv t \la \cds.\frontier$ \Comment{\pp just computed}
  \State $\textsf{found} \la \textsf{false}$
  \For {each $R \in \atoms(Q)$}
  \If {$R.\seekGap(\mv t)$} \Comment{Gap Found}
    \State $\textsf{found} \la \textsf{true}$
    \State $\cds.\insconst(\mv c_R)$ \Comment{the constraint from $R$}
  \EndIf
  \EndFor
  \If {not $\textsf{found}$}
    \State Output $\mv t$
    \State $\cds.\setFrontier(t_0,\dots,t_{n-2},t_{n-1}+1)$
    \EndIf
\EndWhile
\end{algorithmic}
\end{algorithm}

\subsection{\getPP for $\beta$-acyclic queries}

For the sake of clarity, this section presents the \getPP algorithm for 
$\beta$-acyclic queries.
Then, in the next section we show how it is (easily) adapted for general
queries.

Let $\mv p = \langle p_1, \dots, p_k\rangle$ be a pattern. Then, a 
{\em specialization} of $\mv p$ is another pattern $\mv p' = \langle p'_1,
\dots, p'_k\rangle$ {\em of the same length} 
for which $p'_i = p_i$ whenever $p_i \in \mathbb N$. 
In other words, we can get a specialization of $\mv p$ by changing some of the $*$ 
components into equality components.
If $\mv p'$ is a specialization of $\mv p$, then $\mv p$ is a 
{\em generalization} of $\mv p'$.
For two nodes $u$ and $v$ of the \cds, if $P(u)$ is a specialization of $P(v)$, 
then we also say that node $u$ is a specialization of node $v$.

The specialization relation defines a partially ordered set (poset).  When
$\mv p'$ is a specialization of $\mv p$, we write $\mv p' \preceq \mv
p$.  If in addition we know $\mv p' \neq \mv p$, then we write $\mv p'
\prec \mv p$. 

Fix $\frontier = (t_0, t_1,\dots,t_{n-1})$ to be the current frontier.
For $i \in \{0,\dots,n-1\}$, let $G_i$ be the {\em principal filter} generated by
$(t_0,\dots,t_{i-1})$ in this partial order, i.e., it is the set of all nodes $u$ 
(at depth $i$) of the \cds such that $P(u)$ is a generalization of the pattern
$\langle t_0, \dots, t_{i-1} \rangle$ and that $u.\intv \neq \emptyset$.  
Note that $G_0$ consists of only the root node of the \cds.
The key property that we exploit is summarized by the following proposition.

\begin{prop}[From \cite{nnrr}]
Using the notation above, for a $\beta$-acyclic query, there exists a
GAO such that the principal filter $G_i$ generated by any
tuple $(t_0,\dots,t_{i-1})$ is a {\em chain}.
This GAO, called the {\em nested elimination order} (NEO) can be computed
in time linear in the query size.
\label{prop:chain}
\end{prop}

Recall that a chain is a totally ordered set. In particular, for every
$i$, $G = G_i$ has a smallest pattern $\bar{\mv p}$ (or bottom
pattern). 
Thinking of the constraints geometrically, this condition means that the
constraints form a collection of axis-aligned affine subspaces of
$\outspace$ where one is contained inside another.

\begin{algorithm}[!thp]
\caption{$\cds.\getpp()$}
\label{algo:getProbePoint}
\begin{algorithmic}[1]
\Require{$\frontier = (t_0,\dots,t_{n-1})$ is the current frontier}
\Require{$\depth$ is a data member of \cds}
\Ensure{Return \textsc{true} iff a new \pp $\geq \frontier$ is found}
\Ensure{New \pp is stored in $\frontier$}
\Statex

\State $\depth \gets 0$
\State $G_0 \gets \{\textsc{root}\}$
\While {\textsc{true}}
\State $x \la t_{\depth}$
\State $[t_\depth, \isFull] \gets \getFreeValue(x,G_\depth)$
	\If {$\isFull$} 
        \If {$\depth < 0$}
			\State \Return \textsc{false}
		\EndIf
		\State {\bf Continue}
	\EndIf
   \If {$\depth = n-1$}
      \State \Return \textsc{true} \Comment{new \pp computed}
	\EndIf
   \State Compute $G_{\depth+1}$ \Comment{generated by $(t_0,\dots,t_{\depth})$}
   \If {$G_{\depth+1} = \emptyset$ {\bf or} $t_\depth > x$}
      \State $t_i \la -1$  \ \ for all $i \in \{\depth+1,\dots,n-1\}$
   \EndIf
   \If {$G_{\depth+1} = \emptyset$}
      \State \Return {\sc true}
   \EndIf
   \State $\depth \gets \depth+1$
\EndWhile

\end{algorithmic}
\end{algorithm}

Algorithm~\ref{algo:getProbePoint} shows how $\cds$ can return a new
\pp $\geq \frontier$, the current frontier.
For each $\depth$ from $0$ to $n-1$, the algorithm attempts to see if
the current prefix $(t_0,\dots,t_\depth)$ has violated any constraint yet, 
assuming the prefix $(t_0,\dots,t_{\depth-1})$ was already verified to be
a good prefix. Then a new $t_\depth$ value is computed using 
the function $\getFreeValue()$,
which returns the smallest number $\geq t_\depth$ so that the prefix
$(t_0,\dots,t_\depth)$ is good. This new value of $t_\depth$ is completely
determined by the nodes in the set $G_\depth$.

\begin{algorithm}[!thp]
\caption{$\cds.\getFreeValue(x,G)$, where $G$ is a chain}
\label{algo:getFreeValue}
\begin{algorithmic}[1]
\Require{A chain $G$ of nodes, and a starting value $x$}
\Require{$\depth$ is a data member of \cds}
\Require{$\frontier = (t_0,\dots, t_{n-1})$ is a data member of \cds}
\Ensure{Return a pair $[y, \isFull]$, where $y$ is the smallest value 
$\geq x$ not covered by any interval in $v.\intv$, for all $v \in G$, and
$\isFull$ is \textsc{true} if any node in $G$ contains the interval $(-\infty,+\infty)$}
\Statex

\State $\isFull = \textsc{false}$
\If {$G = \emptyset$}
	\State Return $[x, \textsc{false}]$
\EndIf
\State Let $u$ be the bottom node in $G$
\State $y \gets x$
\Repeat
	\State $y \gets u.\nxt(y)$
	\State $[z, \isFull] \gets \getFreeValue(y,G-\{u\})$
	\If {$\isFull$}
		\State Return $[+\infty, \isFull]$
	\EndIf
\Until {$y = z$}
\State $u.\intv.\ins(x-1,y)$ \Comment{Cache!}
\If {$u.\hasNoFreeValue()$}
	\State $\isFull = \textsc{true}$
	\State $\cds$.\truncate($u$)
\ElsIf {$y = +\infty$}
   \State $\depth \la \depth-1$
   \If {$\depth \geq 0$}
      \State $t_{\depth} \la t_{\depth}+1$
   \EndIf
	\State $\isFull = \textsc{true}$
\EndIf
\State Return $[y,\isFull]$
\end{algorithmic}
\end{algorithm}

\begin{idea}[Backtracking and truncating]
The major work is done in $\getFreeValue$
(Algorithm~\ref{algo:getFreeValue}), which ping-pongs among the
interval lists of nodes in the chain $G_\depth$ until a free value is
found. Intervals are inserted to nodes lower in the pecking order to
cache the computation.  If a node in $G_\depth$ is found to contain
the interval $(-\infty,+\infty)$, ruling out {\em all} possible free
values, or if the next free value $t_{\depth}$ is $+\infty$, then the
algorithm has to backtrack.  Algorithm~\ref{algo:fullNode} describes
how to handle a node with no free value.  It finds the first
non-wildcard branch in the \cds from that node to the root, and
inserts an interval to rule out that branch, making sure we never go
down this path again.
\end{idea}

\begin{idea}[Complete node]\label{idea:complete node}
    Repeated calls to $\getFreeValue(x,G_\depth)$ is a major time sink as the
    algorithm ``ping-pongs'' between nodes in $G_\depth$.
    Let $v$ be the bottom node at $G_\depth$.
    Because we start from $-1$, when $t_{\depth}$ reaches $+\infty$ we can 
    infer that $t_\depth$ has iterated through all of the available free 
    values corresponding to the poset $G_\depth$, and $v.\pointlist$ contains 
    all of those free values.
    Hence, at that point we can make node $v$ {\em complete} and the next time
    around when $v$ is the bottom node again we can simply iterate through the
    values in $v.\pointlist$ without ever having to call $\getFreeValue$ on
    $G_\depth$.

    This observation has one caveat: we cannot mark $v.\pointlist$ as complete
    the first time $t_\depth$ reaches $+\infty$ because existing intervals from
    another node in $G_\depth$ might have allowed us to skip inserting some
    intervals into $v.\intv$ the first rotation around those free values.
    The second time $t_\depth$ reaches $+\infty$, however, is safe for inferring
    that $v.\pointlist$ now contains all free values (except for $-\infty,
    +\infty$). Once $v$ is complete, we can iterate through the sorted 
    list $v.\pointlist$, wasting only $O(1)$ amortized time per call.

    The idea of using a complete node is also crucial in speeding up \#\ms as
    explained below. 
    The speedup when this idea is implement is presented in 
    Table~\ref{table:seek gap and complete node}.
Table~\ref{table:seek gap}.
\end{idea}

\begin{table*}[!th]
\centerline{
\begin{tabular}{||l||r|r|r|r|r|r|r|r|r|r|r|r||}
\hline
\hline
 &\begin{sideways}{ca-GrQc}\end{sideways} &\begin{sideways}p2p-Gnutella04\end{sideways} &\begin{sideways}facebook\end{sideways} &\begin{sideways}ca-CondMat\end{sideways} &\begin{sideways}wiki-vote\end{sideways} &\begin{sideways}p2p-Gnutella31\end{sideways} &\begin{sideways}email-Enron\end{sideways} &\begin{sideways}loc-brightkite\end{sideways} &\begin{sideways}soc-Epinions1\end{sideways} &\begin{sideways}soc-Slashdot0811\end{sideways} &\begin{sideways}soc-Slashdot0902\end{sideways} &\begin{sideways}twitter-combined\end{sideways}\\
\hline
\hline
2-comb&1.51&1.48&2.60&1.69&3.46&1.35&3.30&2.98&3.36&4.16&4.11&4.49\\
3-path&1.10&1.42&2.13&1.50&2.82&1.27&2.50&1.91&3.13&3.79&3.64&4.52\\
4-path&1.74&1.49&2.56&2.15&3.34&1.43&3.58&3.26&3.08&3.97&3.93&5.18\\
\hline
\hline
\end{tabular}
}
\label{table:seek gap and complete node}
\caption{Speedup ratio when Idea~\ref{idea:seek gap} and \ref{idea:complete node} are 
incorporated, selectivity is $10$}
\end{table*}

\subsection{\getPP for {\em $\beta$-cyclic} queries} 

When the query is not $\beta$-acyclic, we can no longer infer that the posets
$G_\depth$ are chains. If we follow the algorithm from \cite{nnrr}, then we will
have to compute a transitive closure $\bar G_\depth$ of this poset, and starting
from the bottom node $v$ of $\bar G_\depth$ we will have to recursively poll into
each of the nodes right above $v$ to see if the current value $x$ is free.
While in \cite{nnrr} we showed that this algorithm has a runtime of
$|\cert|^{w+1}$ where $w$ is the elimination width of the GAO, its runtime in
practice is very bad, both due to the large exponent and due to the fact that
specialization branches have to be inserted into the \cds to cache the
computation. 

\begin{table*}[ht]
\centerline{
\renewcommand\tabcolsep{5pt}
\def\arraystretch{1.2}
\begin{tabular}{||l||r|r|r|r|r|r|r|r|r|r|r|r||}
\hline
\hline
 &\begin{sideways}{ca-GrQc}\end{sideways} &\begin{sideways}p2p-Gnutella04\end{sideways} &\begin{sideways}facebook\end{sideways} &\begin{sideways}ca-CondMat\end{sideways} &\begin{sideways}wiki-vote\end{sideways} &\begin{sideways}p2p-Gnutella31\end{sideways} &\begin{sideways}email-Enron\end{sideways} &\begin{sideways}loc-brightkite\end{sideways} &\begin{sideways}soc-Epinions1\end{sideways} &\begin{sideways}soc-Slashdot0811\end{sideways} &\begin{sideways}soc-Slashdot0902\end{sideways} &\begin{sideways}twitter-combined\end{sideways}\\
\hline
\hline
3clique&6.65&31.11&3.68&39.82&3.65&133.38&34.56&72.14&26.66&49.84&53.78&46.90\\
4-clique&8.86&2458.18&3.79&212.02&4.97&1557.35&13.67&188.16&71.70&1736.12&2089.71&$\infty$\\
4-cycle&17.21&387.62&17.34&110.51&4.67&10405.89&558.93&4578.30&5.55&$\infty$&$\infty$&$\infty$\\
\hline
\hline
\end{tabular}
}
\label{table:skipping gap}
\caption{Speedup ratio when Idea~\ref{idea:skipping gap} is
  incorporated. $\infty$ means thrashing.}
\end{table*}

\begin{idea}[Skipping gaps]\label{idea:skipping gap}
     Our idea of speeding up \ms when the query is not $\beta$-acyclic
     is very simple: we compute a $\beta$-cyclic skeleton of the queries, formed
     by a subset of input relations. When we call $\seekGap$ on a relation in
     the $\beta$-acyclic skeleton, the gap box is inserted into the \cds as
     usual. If $\seekGap$ is called on a relation {\em not} in the skeleton, the
     new gap is only used to advance the current frontier $\frontier$ but no new
     constraint is inserted into the \cds.

     Applying this idea risks polling the same gap from a relation (not in the
     $\beta$-acyclic) skeleton more than once. However, we gain time 
     by advancing the frontier and not having to specialize cached intervals 
     into too many branches.
     The speedup when this idea is implement is presented in 
     Table~\ref{table:skipping gap}.
\end{idea}

\subsection{Selecting the GAO}

In theory, \ms \cite{nnrr} requires the GAO to be in {\em nested elimination
order} (NEO). In our implementation, given a query we compute the GAO which
is the NEO with the longest path length. This choice is experimentally
justified in Table~\ref{table:neo}. As evident from the table, the NEO GAOs
yield {\em much} better runtimes than the non-NEO GAOS. Furthermore, the NEO
with the longer path (ABCDE) is better than the others NEO GAOs because the
longer paths allow for more caching to take effect. There are $5!$ different
attribute orderings for the $4$-path queries but many of them are isomorphic;
that's why we have presented only $7$ representative orderings.

\begin{table*}[!th]
\centerline{
\renewcommand\tabcolsep{5pt}
\def\arraystretch{1.2}
\begin{tabular}{||l||r|r|r|r|r||r|r|r||}
\hline
&\multicolumn{5}{c|}{NEO GAOs}&\multicolumn{2}{c|}{non-NEO GAOS}&\\
\hline
Data set&ABCDE&BACDE&BCADE&CBADE&CBDAE&ABDCE&BADCE&Edge set sizes\\
\hline
\hline
ca-GrQc&0.10&   0.12&   0.11&   0.12&   0.12&   0.94&   1.00&   14484\\
p2p-Gnutella04& 0.24 &  0.25&   0.23&   0.22&   0.23&   7.14 &  7.05 &  39994\\
facebook&   9.99&   10.86&  12.78 & 14.80  &15.58  &30.28 & 25.18 & 88234\\
ca-CondMat& 0.37 &  0.57&   0.58&   0.54&   0.53&   20.47&  21.72 & 93439\\
wiki-vote&  16.99 & 30.17  &39.01  &34.32 & 34.74&  61.22 & 61.47 & 100762\\
p2p-Gnutella31& 0.67 &  0.64&   0.69&   0.64&   0.67&   81.69&  85.19 & 147892\\
email-Enron&14.33 & 20.84&  25.72&  19.51&  19.75 & 81.09 & 83.05&  183831\\
loc-brightkite& 4.45 &  5.48&   6.73&   5.28&   4.39&   105.38& 107.23 &214078\\
\hline
\hline
\end{tabular}
}
\caption{Runtimes of \ms on $4$-path query with different GAOs}
\label{table:neo}
\end{table*}

\subsection{Multi-threading implementation}

To parallelize \ms, our strategy is very simple: we partition the output space
into $p$ equal-sized parts, where $p$ is determined by the number of CPUs times
a {\em granularity factor} $f$, where $f=1$ for acyclic queries and
$f=8$ for cyclic queries. These values are determined after minor ``micro
experiments'' to be shown below.
Each part represents a job submitted to a {\em job pool}, a facility supported
by LogicBlox' engine. 
We set $f>1$ for cyclic queries because the parts are not born
equal: some threads might be finishing much earlier than other threads, and it
can go grab the next unclaimed job from the pool; this is a form of
{\em work stealing}. 

We do not set $f$ to be too large because there is a diminishing return point
after which the overhead of having too many threads dominates the work stealing
saving. On the other hand, setting $f$ to be larger also helps prevents 
thrashing in case the input is too large. Each thread can release the memory
used by its \cds before claiming the next job.

Table~\ref{table:work stealing} shows the average {\em normalized}
runtimes over varying granularity factor $f$. Here, the runtimes are
divided by the runtime of \ms when $f=1$.

\begin{table}[ht]
\centerline{
\def\arraystretch{1.2}
\begin{tabular}{||l||r|r|r|r|r|r|r||}
\hline
\hline
Granularity&1&2&3&4&8&12&14\\
\hline
\hline
3-path&1&   0.97&   1.04&   1.12&   1.37&   1.55&   1.65\\
4-path&1 &  0.92 &  0.91 &  0.99 &  0.96 &  0.98 &  0.98\\
2-comb&1 &  0.90&   0.94&   0.96&   1.09&   1.21&   1.26\\
3-clique&1& 0.88&   0.89&   0.92&   0.98&   1.07&   1.09\\
4-clique&1& 0.91 &  0.82 &  0.82 &  0.82 &  0.86 &  0.87\\
4-cycle&1 & 0.84 &  0.84 &  0.83 &  0.87 &  0.91 &  0.92\\
\hline
\hline
\end{tabular}
}
\caption{Average normalized runtime across partition granularity}
\label{table:work stealing}
\end{table}

\subsection{\#\ms}

\begin{idea}[Micro message passing]\label{idea:message passing}
    When a node is complete (Idea~\ref{idea:complete node}), we know that the
    points in its $\pointlist$ (except for $-\infty$ and $+\infty$) are the
    start of branches down the search space that have already been computed.
    For example, consider the query 
    \[ Q = R_1(A,B) \Join R_2(A, C) \Join R_3(B, D) \Join R_4(C) \Join R_5(D),
    \]
    where the GAO is $A, B, C, D$.
    At $\depth=2$, corresponding to intervals on attribute $C$, the bottom
    node of $G_2$ might be $\dbox{a, *}$ for some $a \in \Dom$.
    When this node is complete, we know that the join
    $R_4(C) \Join \pi_C \sigma_{A=a} R_1$ is already computed and the output
    points are stored in $\dbox{a,*}.\pointlist$. Hence, the size of the
    $\pointlist$ (minus $2$) is the size of this join, which should be {\em
    multiplied} with the number of results obtained from the independent branch
    of the search space: $\sigma_{A=a} R_1 \Join R_3(B, D) \Join R_5(D)$.

    Our idea here is to keep a count value associated with each point in the
    $\pointlist$ at each node. When a node is completed for the first time,
    it sums up all counts in its $\pointlist$, traces back the \cds to find the
    first equality branch, and multiply this sum with the corresponding count.
    For example, to continue with the above example, when node $\dbox{a}$ is 
    complete, it will take the sum over all count values of the points $b$ in
    $\dbox{a}.\pointlist$, and multiply the result with the count of the point
    $a$ in $\mathsf{root}.\pointlist$. (Initially, all count values are $1$.)
    The tally from $\dbox{a,*}$ was multiplied there too.
    In particular, if a node is already completed, we no longer have to iterate
    through the points in its $\pointlist$.
    \#\ms is to message passing what \ms was to Yannakakis algorithm: \#\ms
    does not pass large messages, only the absolutely necessary counts are sent
    back up the tree.
\end{idea}

\subsection{Lollipop queries and a hybrid algorithm}
\label{subsec:lollipop}

The $2$-lollipop query is a $2$-path followed by a $3$-clique:
$$(A)(AB)(BC)(CD)(DE)(CE).$$
The $3$-lollipop query is a $3$-path followed by a $4$-clique, in the same
manner.
To illustrate that a combination of \ms and \lftj ideas might be ideal,
we crafted a specialized algorithm that runs \ms on the path part of the
query and \lftj on the clique part. In particular, this {\em hybrid algorithm}
allows for Idea~\ref{idea:complete node} to be applied on attributes
$A$, $B$, and $C$ of the query, and for Idea~\ref{idea:skipping gap} to be
implemented completely on the clique part of the query: all gaps are used to
advance the frontier.

\begin{algorithm}
\caption{$\cds$.\truncate($u$)}
\label{algo:fullNode}
\begin{algorithmic}[1]
\Require{$\depth$ is a data member of \cds}
\Ensure{Truncate the \cds to cut $u$ off and update $\depth$.}
\Statex
	\While {$\depth \geq 0$}
		\State $\depth \gets \depth-1$
		\If {$\depth < 0$}
			\State Return
		\EndIf
		\State Let $v$ be a parent node of $u$ in the \cds
		\If {$vu$ is labeled with a value $x \in \Dom$}
			\State $v.\intv.\ins(x-1,x+1)$
			\State \Return
		\Else
			\State $u \gets v$
		\EndIf
	\EndWhile
\end{algorithmic}
\end{algorithm}
\section{Experiments}


\subsection{Data sets, Queries, and Setup}

The data sets we use for our experiments come from the SNAP network
data sets collection~\cite{snapnets}. We use the following data sets:

\begin{center}
\vskip1em
{\sf\small
\def\arraystretch{1.2}
\begin{tabular}{l|r|r|r}
  name & nodes & edges & triangle count\\
  \hline
  wiki-Vote         &  7,115 & 103,689  &   608,389 \\
  p2p-Gnutella31    & 62,586 &  147,892 &     2,024 \\
  p2p-Gnutella04    & 10,876 & 39,994   &       934 \\
  loc-Brightkite    & 58,228 & 428,156  &   494,728 \\
  ego-Facebook      &  4,039 &  88,234  & 1,612,010 \\
  email-Enron       & 36,692 & 367,662  &   727,044 \\
  ca-GrQc           &  5,242 &  28,980  &    48,260 \\
  ca-CondMat        & 23,133 & 186,936  &   173,361 \\
  \hline
  ego-Twitter       & 81,306 & 2,420,766 & 13,082,506 \\
  soc-Slashdot0902  & 82,168 &   948,464 &    602,592 \\
  soc-Slashdot0811  & 77,360 &   905,468 &    551,724 \\
  soc-Epinions1     & 75,879 &   508,837 &  1,624,481 \\
  \hline
  soc-Pokec         & 1,632,803 &  30,622,564 &  32,557,458 \\
  soc-LiveJournal1  & 4,847,571 &  68,993,773 & 285,730,264 \\
  com-Orkut         & 3,072,441 & 117,185,083 & 627,584,181 \\
\end{tabular}
}
\vskip1em
\end{center}

Some queries also require a subset of nodes to be used as part of the
queries. We execute these queries with different random samples of
nodes, with varying size. A random sample of nodes is created by
selecting nodes with probably $1/s$, where $s$ is referred to as
selectivity in our results. For example, for selectivity 10 and 100 we
select respectively approximately 10\% and 1\% of the nodes.

\paragraph*{Queries}

We execute experiments with the following queries. We include the
Datalog formulation. Variants for other systems (e.g. SQL, SPARQL) are
available online.
\begin{itemize}\compactify
\item $\{3,4\}$-clique: find subgraphs with $\{3,4\}$ nodes such that
  every two nodes are connected by an edge. The 3-clique query is also
  known as the triangle problem. Similar to other work, we treat
  graphs as undirected for this query.

  \texttt{\small edge(a,b), edge(b,c), edge(a,c), a<b<c.}

\item $4$-cycle: find cycles of length 4.

  \texttt{\small edge(a,b), edge(b,c), edge(c,d), edge(a,d), a<b<c<d}

\item $\{3,4\}$-path: find paths of length $\{3,4\}$ for all
  combinations of nodes \texttt{a} and \texttt{b} from two random
  samples \texttt{v1} and \texttt{v2}.

  \texttt{\small v1(a), v2(d), edge(a, b), edge(b, c), edge(c, d).}

\item $\{1,2\}$-tree: find complete binary trees with $2n$ leaf nodes
  s.t. each leaf node is drawn from a different random sample.

  \texttt{\small v1(b), v2(c), edge(a, b), edge(a, c).}

\item $2$-comb: find left-deep binary trees with $2$ leaf nodes s.t.
  each leaf node is drawn from a different random sample.

  \texttt{\small v1(c), v2(d), edge(a, b), edge(a, c), edge(b, d).}

\item $\{2,3\}$-lollipop: finds $\{2,3\}$-path subgraphs followed by
  $\{3,4\}$-cliques, as described in \ref{subsec:lollipop}. The start
  nodes `\texttt{a}' are a random sample `\texttt{v1}'.

  \texttt{\small v1(a), edge(a, b), edge(b, c), edge(c, d), edge(d, e), edge(c, e).}
\end{itemize}

The queries can be divided in acyclic and cyclic queries. This
distinction is important because Minesweeper is instance-optimal for
the acyclic queries~\cite{nnrr}. From our queries, $n$-clique and
$n$-cycle are $\beta$-cyclic. All others are $\beta$-acyclic. We add
predicates $v_1$ and $v_2$. As we vary the size of these predicates,
we also change the amount of redundant work. Minesweeper is able to
exploit this redundancy, as we show below. All queries are executed as
a count, which returns the number of results to the client. We
verified the result for all implementations.

\paragraph*{Systems}
We evaluate the performance of LogicBlox using Minesweeper and LFTJ by
comparing the performance of a wide range of database systems and
graph engines.

\vskip1em
{\sf\small
\def\arraystretch{1.2}
\begin{tabular}{l|l}
  name & description\\
  \hline
  lb/lftj   & LogicBlox 4.1.4 using LFTJ\\
  lb/ms     & LogicBlox 4.1.4 using Minesweeper\\
  psql      & PostgreSQL 9.3.4\\
  monetdb   & MonetDB 1.7 (Jan2014-SP3)\\
  virtuoso  & Virtuoso 7\\
  neo4j     & Neo4j 2.1.5\\
  graphlab  & GraphLab v2.2\\
\end{tabular}
}
\vskip1em

We select such a broad range of systems because the performance of
join algorithms is not primarily related to the storage architecture
of a database (e.g. row vs column vs graph stores). Also, we want to
evaluate whether general-purpose relational databases utilizing
optimal join algorithms can replace specialized systems, like graph
databases, and perhaps even graph engines.

Due to the complexity of implementing and tuning the queries across all
these systems (e.g.  tuning the query or selecting the right indices),
we first select two queries that we execute across the full range of
systems. After establishing that we can select representative systems
without compromising the validity of our results, we run the remaining
experiments across the two variants of LogicBlox, PostgreSQL, and
MonetDB. The results will show that the graph databases have their
performance dominated by our selected set. We evaluate GraphLab only
for $3$-clique and $4$-clique queries. The $3$-clique implementation
is included in the GraphLab distribution and used as-is. We developed
the $4$-clique implementation with advice from the GraphLab community,
but developing new algorithms on GraphLab can be a heavy undertaking,
requiring writing C++ and full understanding of its imperative {\em
  gather-apply-scatter} programming model. Therefore, we cannot
confidently extend coverage on GraphLab beyond these queries.

\newcommand{\dataset}[1]{\begin{sideways}#1\end{sideways}}

\newcommand{\best}[1]{\textbf{#1}}

\begin{table*}[ht]
\begin{center}%
{\sf\small%
\def\arraystretch{1.2}%
\renewcommand\tabcolsep{6pt}%
\begin{tabular}{l|l|l|rrrrrrrr|rrrr|rrr}
  \multicolumn{3}{l|}{}
  & \dataset{wiki-Vote}
  & \dataset{p2p-Gnutella31}
  & \dataset{p2p-Gnutella04}
  & \dataset{loc-Brightkite}
  & \dataset{ego-Facebook}
  & \dataset{email-Enron}
  & \dataset{ca-GrQc}
  & \dataset{ca-CondMat}
  & \dataset{ego-Twitter}
  & \dataset{soc-Slashdot0902}
  & \dataset{soc-Slashdot0811}
  & \dataset{soc-Epinions1}
  & \dataset{soc-Pokec}
  & \dataset{soc-LiveJournal1}
  & \dataset{com-Orkut}
  \\

\hline
\multicolumn{2}{l}{3-clique} & lb/lftj & 0 & 0 & 0 & 0 & 0 & 0 & 0 & 0 & 5 & 1 & 1 & 1 & 75 & 165 & 742 \\ \cline{4-18}
\multicolumn{2}{l}{} & lb/ms & 1 & 1 & 0 & 2 & 1 & 3 & 0 & 1 & 23 & 7 & 5 & 6 & 282 & - & - \\ \cline{4-18}
\multicolumn{2}{l}{} & psql & 1446 & 6 & 2 & - & 575 & - & 10 & 348 & - & - & - & - & - & - & - \\ \cline{4-18}
\multicolumn{2}{l}{} & monetdb & - & 3 & 3 & 945 & 947 & - & 22 & 98 & - & - & - & - & - & - & - \\ \cline{4-18}
\multicolumn{2}{l}{} & virtuoso & 18 & 2 & 1 & 17 & 23 & 46 & 1 & 4 & 296 & 75 & 68 & 158 & - & - & - \\ \cline{4-18}
\multicolumn{2}{l}{} & neo4j & 348 & 19 & 6 & 212 & 250 & 418 & 4 & 32 & - & 1441 & 1308 & 1745 & - & - & - \\ \cline{4-18}
\multicolumn{2}{l}{} & graphlab & 0 & 0 & 0 & 0 & 0 & 0 & 0 & 0 & 0 & 0 & 0 & 0 & 3 & 7 & 27 \\

\hline
\multicolumn{2}{l}{4-clique} & lb/lftj & 3 & 0 & 0 & 11 & 9 & 4 & 0 & 1 & 427 & 4 & 4 & 13 & 644 & - & - \\ \cline{4-18}
\multicolumn{2}{l}{} & lb/ms & 11 & 1 & 0 & 10 & 31 & 25 & 1 & 2 & 288 & 39 & 32 & 96 & - & - & - \\ \cline{4-18}
\multicolumn{2}{l}{} & psql & - & 52 & 10 & - & - & - & 1021 & - & - & - & - & - & - & - & - \\ \cline{4-18}
\multicolumn{2}{l}{} & monetdb &  & 17 & 15 &  &  &  & 1219 & - & - & - & - & - & - & - & - \\ \cline{4-18}
\multicolumn{2}{l}{} & virtuoso & 447 & 2 & 0 & 364 & 1240 & 968 & 2 & 38 & - & 1427 & 1273 & - & - & - & - \\ \cline{4-18}
\multicolumn{2}{l}{} & neo4j & - & - & - & - & - & - & - & - & - & - & - & - & - & - & - \\ \cline{4-18}
\multicolumn{2}{l}{} & graphlab & 0 & 0 & 0 & 0 & 1 & 0 & 0 & 0 & 6 & 1 & 1 & 1 & - & - & - \\

\hline
\multicolumn{2}{l}{4-cycle} & lb/lftj & 11 & 1 & 0 & 4 & 8 & 7 & 0 & 1 & 171 & 31 & 29 & 34 & 1416 & - & - \\ \cline{4-18}
\multicolumn{2}{l}{} & lb/ms & 24 & 3 & 1 & 17 & 23 & 59 & 0 & 3 & 587 & 183 & 156 & 268 & - & - & - \\ \cline{4-18}
\multicolumn{2}{l}{} & psql & 309 & 4 & 1 & 1394 & 539 & - & 47 & 112 & - & - & - & - & - & - & - \\ \cline{4-18}
\multicolumn{2}{l}{} & monetdb & 502 & 1 & 1 & 657 & 347 & - & 19 & 60 & - & - & - & - & - & - & - \\
\end{tabular}
}

\caption{Duration of cyclic queries \{3,4\}-clique and 4-cycle in seconds. ``-'' denotes a timeout.}
\label{fig:results-cyclic}
\end{center}
\end{table*}

\paragraph*{Hardware}
For all systems, 
we use AWS EC2 m3.2xlarge instances. This instance type has an
Intel Xeon E5-2670 v2 Ivy Bridge or Intel Xeon E5-2670 Sandy Bridge
processor with 8 hyperthreads and 30GB of memory. Database files are
placed on the 80GB SSD drive provided with the instance. We use Ubuntu
14.04 with PostgreSQL from Ubuntu's default repository and the other
systems installed manually. 

\paragraph*{Protocol} We execute each experiment three times
and average the last two executions. We impose a timeout of 30 minutes
(1800 seconds) per execution. For queries that require random samples
of nodes, we execute them with multiple selectivities. For small data
sets we use selectivity 8 (12.5\%) and 80 (1.25\%). For the other data
sets we use selectivities of 10 (10\%), 100 (1\%), and 1000
(0.1\%). We ensure each system sees the same random datasets. Although
across runs for the same system, we use different random draws.

\subsection{Results}

We validate that worst-case optimal algorithms like LFTJ outperform
many systems on cyclic queries, while \ms is fastest on acyclic
queries.

\subsubsection{Standard Benchmark Queries}

Clique finding is a popular benchmark task that is hand optimized by
many systems. Table~\ref{fig:results-cyclic} shows that both 
LFTJ and Minesweeper are faster than all systems except the graph engine GraphLab on
3-clique. On our C++ implementation of 4-clique, GraphLab runs out of
memory for big data-sets. After the systems that implement the optimal
join algorithms, Virtuoso is fastest. Relational systems that do
conventional joins perform very poorly on 3-clique and 4-clique due to
extremely large intermediate results of the self-join, whether
materialized or not. The simultaneous search for cliques as performed
by Minesweeper and LFTJ prevent this. This difference is particularly
striking on 4-clique.

LFTJ and Minesweeper perform well on datasets that have few
cliques. This is visible in the difference between Twitter vs Slashdot
and Epinions data sets in which the performance is much closer to
GraphLab. 

\newcommand*\rot{\rotatebox{315}}

\newcommand{\datasetsmall}[1]{\multicolumn{2}{c|}{\begin{sideways}\ #1\end{sideways}}}
\newcommand{\datasetbig}[1]{\multicolumn{3}{c|}{\begin{sideways}\ #1\end{sideways}}}
\newcommand{\datasetfinal}[1]{\multicolumn{3}{c}{\begin{sideways}\ #1\end{sideways}}}
\newcommand{\query}[1]{#1}
\newcommand{\system}[1]{#1}

\begin{sidewaystable*}[p]
{\sf\scriptsize
\renewcommand\tabcolsep{2.2pt}
\def\arraystretch{1.3}
\begin{tabular}{ll|rr|rr|rr|rr|rr|rr|rr|rr|rrr|rrr|rrr|rrr|rrr|rrr|rrr}

\multicolumn{2}{c|}{}
  & \datasetsmall{wiki-Vote}
  & \datasetsmall{p2p-Gnutella31}
  & \datasetsmall{p2p-Gnutella04}
  & \datasetsmall{loc-Brightkite}
  & \datasetsmall{ego-Facebook}
  & \datasetsmall{email-Enron}
  & \datasetsmall{ca-GrQc}
  & \datasetsmall{ca-CondMat}
  & \datasetbig{ego-Twitter}
  & \datasetbig{soc-Slashdot0902}
  & \datasetbig{soc-Slashdot0811}
  & \datasetbig{soc-Epinions1}
  & \datasetbig{soc-Pokec}
  & \datasetbig{soc-LiveJournal1}
  & \datasetfinal{com-Orkut}
  \\
\multicolumn{2}{c|}{}
  & 80 & 8
  & 80 & 8
  & 80 & 8
  & 80 & 8
  & 80 & 8
  & 80 & 8
  & 80 & 8
  & 80 & 8
  & 1K & 100 & 10
  & 1K & 100 & 10
  & 1K & 100 & 10
  & 1K & 100 & 10
  & 1K & 100 & 10
  & 1K & 100 & 10
  & 1K & 100 & 10
  \\

\hline
\query{3-path}
	& \system{lb/lftj}	 & 0 & 2 & \best{0} & \best{0} & \best{0} & \best{0} & 1 & 20 & 0 & 0 & 1 & 40 & 0 & 0 & 0 & 2 & \best{1} & 13 & 144 & 1 & 8 & 98 & 1 & 8 & 110 & 0 & 5 & 27 & 9 & \best{120} & 1521 & \best{61} & 1035 & - & \best{60} & 825 & - \\
	& \system{lb/ms}	 & 0 & \best{1} & 0 & 0 & 0 & 0 & \best{1} & \best{4} & 0 & \best{0} & \best{1} & \best{3} & 0 & \best{0} & 1 & \best{1} & 1 & \best{5} & \best{18} & 1 & \best{4} & \best{10} & 1 & \best{4} & \best{10} & 0 & \best{1} & \best{4} & 25 & 129 & \best{408} & 68 & \best{259} & - & 111 & \best{451} & - \\
	& \system{psql}	 & \best{0} & 12 & 0 & 0 & 0 & 0 & 2 & 203 & \best{0} & 3 & 3 & 556 & \best{0} & 0 & \best{0} & 7 & 2 & 215 & - & \best{0} & 5 & 938 & \best{0} & 6 & 890 & \best{0} & 2 & 243 & \best{8} & 166 & - & 142 & 1011 & - & - & - & - \\
	& \system{monetdb}	 & 128 & 131 & 1 & 1 & 0 & 0 & 993 & 1036 & 45 & 56 & - & - & 6 & 5 & 57 & 68 & - & - & - & - & - & - & - & - & - & - & - & - & - & - & - & - & - & - & - & - & - \\
	& \system{virtuoso}	 & 1 & 16 & 0 & 0 & 0 & 0 & 18 & 319 & 0 & 4 & 37 & 719 & 0 & 1 & 1 & 10 & 7 & 59 & 1435 & 8 & 52 & 1433 & 6 & 65 & 1268 & 2 & 15 & 403 & 75 & 784 & - & - & - & - & - & - & - \\
	& \system{neo4j}	 & 4 & 71 & 1 & 2 & 0 & 1 & 82 & 633 & 4 & 19 & 163 & 1584 & 1 & 4 & 6 & 42 & 57 & 323 & - & 28 & 370 & - & 41 & 405 & - & 15 & 88 & 877 & - & - & - & - & - & - & - & - & - \\

\hline
\query{4-path}
	& \system{lb/lftj}	 & 4 & 193 & \best{0} & \best{0} & 0 & \best{0} & 44 & 1155 & 1 & 9 & 75 & - & 1 & 5 & 6 & 59 & 103 & 1286 & - & \best{3} & 203 & - & 62 & 240 & - & 4 & 68 & - & 710 & - & - & - & - & - & - & - & - \\
	& \system{lb/ms}	 & \best{1} & \best{1} & 0 & 1 & 0 & 0 & \best{4} & \best{9} & \best{0} & \best{1} & \best{4} & \best{7} & \best{0} & \best{0} & \best{2} & \best{4} & \best{8} & \best{22} & \best{46} & 7 & \best{13} & \best{24} & \best{7} & \best{14} & \best{23} & 2 & \best{6} & \best{10} & 206 & \best{556} & - & \best{470} & - & - & \best{697} & - & - \\
	& \system{psql}	 & 3 & 1099 & 0 & 1 & \best{0} & 0 & 299 & - & 0 & 102 & 914 & - & 0 & 39 & 4 & 437 & - & - & - & 9 & 1211 & - & 10 & 1637 & - & \best{1} & 470 & - & \best{94} & - & - & - & - & - & 1378 & - & - \\
	& \system{monetdb}	 & - & - & 3 & 4 & 1 & 2 & - & - & - & - & - & - & 230 & 321 & - & - & - & - & - & - & - & - & - & - & - & - & - & - & - & - & - & - & - & - & - & - & - \\
	& \system{virtuoso}	 & 30 & 1363 & 0 & 1 & 0 & 0 & 1664 & - & 5 & 189 & - & - & 4 & 29 & 37 & 577 & 710 & - & - & 1058 & - & - & 657 & - & - & 46 & 1785 & - & - & - & - & - & - & - & - & - & - \\
	& \system{neo4j}	 & 161 & - & 1 & 7 & 0 & 3 & - & - & 105 & 437 & - & - & 23 & 109 & 201 & 1309 & - & - & - & - & - & - & - & - & - & 1097 & - & - & - & - & - & - & - & - & - & - & - \\

\hline
\query{1-tree}
	& \system{lb/lftj}	 & 0 & 0 & \best{0} & \best{0} & \best{0} & \best{0} & \best{0} & \best{0} & \best{0} & \best{0} & \best{0} & \best{0} & 0 & 0 & \best{0} & \best{0} & \best{0} & \best{0} & \best{2} & \best{0} & \best{0} & \best{1} & \best{0} & \best{0} & \best{1} & \best{0} & \best{0} & \best{0} & 1 & \best{3} & \best{30} & \best{1} & \best{7} & \best{82} & \best{2} & \best{32} & 443 \\
	& \system{lb/ms}	 & 0 & \best{0} & 0 & 0 & 0 & 0 & 0 & 1 & 0 & 0 & 0 & 0 & 0 & 0 & 0 & 0 & 1 & 2 & 2 & 1 & 1 & 1 & 1 & 1 & 1 & 0 & 1 & 1 & 28 & 32 & 46 & 55 & 64 & 97 & 79 & 100 & \best{152} \\
	& \system{psql}	 & \best{0} & 1 & 0 & 0 & 0 & 0 & 0 & 1 & 0 & 0 & 0 & 2 & \best{0} & \best{0} & 0 & 0 & 0 & 1 & 44 & 0 & 0 & 4 & 0 & 0 & 1 & 0 & 0 & 2 & \best{1} & 17 & 160 & 25 & 36 & 513 & 2 & 106 & - \\
	& \system{monetdb}	 & 4 & 5 & 1 & 1 & 0 & 0 & - & - & - & - & - & - & 0 & 0 & 1 & 1 & 88 & 78 & 95 & - & - & - & - & - & - & 12 & 11 & 10 & - & - & - & - & - & - & - & - & - \\

\hline
\query{2-tree}
	& \system{lb/lftj}	 & 8 & - & 1 & \best{1} & 0 & 1 & 531 & - & 3 & - & - & - & \best{0} & 250 & 6 & - & \best{2} & - & - & \best{1} & - & - & \best{2} & - & - & \best{0} & 560 & - & 587 & - & - & - & - & - & - & - & - \\
	& \system{lb/ms}	 & \best{1} & \best{1} & 1 & 2 & 0 & \best{1} & \best{6} & \best{9} & \best{0} & \best{1} & \best{4} & \best{8} & 0 & \best{1} & \best{3} & \best{4} & 21 & \best{32} & \best{45} & 10 & \best{15} & \best{23} & 9 & \best{15} & \best{22} & 4 & \best{6} & \best{10} & \best{561} & \best{704} & - & \best{977} & \best{1249} & - & \best{1315} & - & - \\
	& \system{psql}	 & - & - & \best{0} & 15 & \best{0} & 6 & - & - & 1622 & - & - & - & 61 & - & 1228 & - & - & - & - & - & - & - & - & - & - & - & - & - & - & - & - & - & - & - & - & - & - \\
	& \system{monetdb}	 & - & - & - & - & - & - & - & - & - & - & - & - & - & - & - & - & - & - & - & - & - & - & - & - & - & - & - & - & - & - & - & - & - & - & - & - & - \\

\hline
\query{2-comb}
	& \system{lb/lftj}	 & 0 & 6 & \best{0} & \best{0} & 0 & \best{0} & 1 & 20 & \best{0} & 3 & 1 & 50 & 0 & 0 & 0 & 2 & \best{1} & 15 & 180 & 1 & 8 & 117 & 1 & 11 & 101 & 0 & 5 & 41 & \best{11} & \best{140} & 1780 & \best{66} & 1161 & - & 395 & - & - \\
	& \system{lb/ms}	 & 0 & \best{0} & 0 & 1 & 0 & 0 & \best{1} & \best{3} & 0 & \best{0} & \best{1} & \best{2} & 0 & \best{0} & 1 & \best{1} & 2 & \best{7} & \best{12} & 1 & \best{4} & \best{6} & 1 & \best{4} & \best{6} & 1 & \best{1} & \best{3} & 64 & 156 & \best{272} & 128 & \best{282} & \best{507} & 312 & \best{575} & - \\
	& \system{psql}	 & \best{0} & 51 & 0 & 0 & \best{0} & 0 & 2 & 206 & 0 & 29 & 3 & 553 & \best{0} & 0 & \best{0} & 6 & 2 & 205 & - & \best{0} & 5 & 1014 & \best{0} & 6 & 936 & \best{0} & 3 & 288 & 14 & 196 & - & 153 & 1111 & - & \best{162} & - & - \\
	& \system{monetdb}	 & 388 & 478 & 3 & 3 & 1 & 1 & - & - & - & - & - & - & 5 & 5 & 53 & 62 & - & - & - & - & - & - & - & - & - & - & - & - & - & - & - & - & - & - & - & - & - \\

\hline
\query{2-lollipop}
	& \system{lb/lftj}	 & 7 & 189 & 0 & \best{0} & 0 & 0 & 144 & - & 9 & 14 & 468 & - & 2 & 4 & 6 & 36 & 185 & 829 & - & 88 & 664 & - & 130 & 671 & - & 77 & 235 & - & 396 & - & - & - & - & - & - & - & - \\
	& \system{lb/ms}	 & 16 & 169 & 0 & 1 & 0 & 0 & 407 & - & 25 & 38 & - & - & 5 & 12 & 18 & 73 & 517 & - & - & 230 & 1498 & - & 233 & - & - & 167 & 439 & - & - & - & - & - & - & - & - & - & - \\
	& \system{lb/hybrid}	 & \best{1} & \best{1} & \best{0} & 0 & \best{0} & \best{0} & \best{7} & \best{8} & \best{0} & \best{1} & \best{10} & \best{13} & \best{0} & \best{0} & \best{1} & \best{2} & \best{18} & \best{37} & \best{58} & \best{17} & \best{26} & \best{30} & \best{26} & \best{47} & \best{51} & \best{8} & \best{13} & \best{15} & \best{203} & \best{625} & \best{878} & \best{1080} & - & - & \best{1663} & - & - \\
	& \system{psql}	 & 286 & - & 0 & 3 & 0 & 1 & - & - & 209 & 724 & - & - & 20 & 146 & 356 & - & - & - & - & - & - & - & - & - & - & - & - & - & - & - & - & - & - & - & - & - & - \\
	& \system{monetdb}	 & 92 & - & 0 & 1 & 0 & 0 & - & - & 41 & 203 & - & - & 10 & 50 & 93 & 947 & - & - & - & - & - & - & - & - & - & 1208 & - & - & - & - & - & - & - & - & - & - & - \\

\hline
\query{3-lollipop}
	& \system{lb/lftj}	 & - & - & \best{0} & 1 & 0 & 1 & - & - & - & - & - & - & - & - & - & - & - & - & - & - & - & - & - & - & - & - & - & - & - & - & - & - & - & - & - & - & - \\
	& \system{lb/ms}	 & - & - & 1 & 3 & 0 & 1 & - & - & - & - & - & - & - & - & - & - & - & - & - & - & - & - & - & - & - & - & - & - & - & - & - & - & - & - & - & - & - \\
	& \system{lb/hybrid}	 & \best{19} & \best{20} & 0 & \best{1} & \best{0} & \best{0} & \best{193} & \best{195} & \best{21} & \best{26} & \best{313} & \best{312} & \best{6} & \best{8} & \best{25} & \best{27} & \best{1680} & - & - & \best{477} & \best{485} & \best{483} & \best{642} & \best{650} & \best{1263} & \best{255} & \best{275} & \best{281} & - & - & - & - & - & - & - & - & - \\
	& \system{psql}	 & - & - & 4 & 35 & 3 & 25 & - & - & - & - & - & - & - & - & - & - & - & - & - & - & - & - & - & - & - & - & - & - & - & - & - & - & - & - & - & - & - \\
	& \system{monetdb}	 & - & - & 1 & 18 & 1 & 11 & - & - & - & - & - & - & - & - & - & - & - & - & - & - & - & - & - & - & - & - & - & - & - & - & - & - & - & - & - & - & - \\

\hline
\end{tabular}
}
\caption{Duration (seconds) of acyclic queries with different
  selectivities (80 and 8 for small datasets, 1K, 100 and 10 for
  others). ``-'' denotes a timeout.}
\label{fig:results-acyclic}
\end{sidewaystable*}

\paragraph*{Acyclic Queries: \{3,4\}-path}
Table~\ref{fig:results-acyclic} shows the results for \{3,4\}-path and
other acyclic queries. Minesweeper is faster here,
outperforming LFTJ on virtually every data set for 3-path. 

\ms does very well for non-trivial acyclic queries such as
\{3,4\}-path queries because it has a caching mechanism that enables
it to prune branches using the \cds. Interestingly, PostgreSQL is now
the next fastest system: it is even more efficient than the worst-case
optimal join system for a few data sets on 3-path. The PostgreSQL
query optimizer is smart enough to determine that it is best to start
separately from the two node samples, and materialize the intermediate
result of one of the edge subsets ($\texttt{v1}(a) \Join
\texttt{edge}(a, b)$ or $\texttt{v2}(d) \Join \texttt{edge}(c,
d)$). This strategy starts failing though on 4-path, due to two edge
joins between these two results, as opposed to just one for
3-path. MonetDB starts from either of the random node samples, and
immediately does a self-join between two edges, which is a slow
execution plan.

LFTJ does relatively worse on 4-path, and times out on bigger
datasets. LFTJ with variable ordering $[a,b,d,c]$ is fairly similar to
a nested loop join where for every $\texttt{v1}(a) \Join
\texttt{edge}(a, b)$ the join $\texttt{v2}(d) \Join \texttt{edge}(c,
d)$ is computed, except that the last join includes a filter on $
\texttt{edge}(b, c)$ for the current $b$. This is still workable for
3-path, but does not scale to 4-path for bigger data sets. The
comparison with \{3,4\}-clique and 4-cycle is interesting here,
because the join is very similar. These queries allow LFTJ to evaluate
the self-joins from both directions, where one direction narrows down
the search of the other. This is not applicable to \{3,4\}-path. This
example shows that LFTJ does not eliminate the need for the query
optimizer to make smart materialization decisions for some joins. If
part of the 3-path join is manually materialized, then performance
improves.

For 3-path, \ms and LFTJ have an interesting difference in performance
when executing with different sizes of random node samples. LFTJ is
consistently the fastest of the two algorithms for very high
selectivity, but Minesweeper is best with lower selectivity, where
Minesweeper starts benefiting substantially from the
caching mechanism. With lower selectivity, the amount of redundant work is increased
due to repeatedly searching for sub-paths. To deal with this type of queries, we need to have
a mechanism to not only be able to do simultaneous search, which both LTFT and Minesweeper have and 
perform well on clique-type queries, but also to avoid any redundant work generated when computing
the sub-graphs. The latter is easily integrated into Minesweeper and that integration is very natural
as we can see in Section~\ref{sec:minesweeper}.
Also in Section~\ref{extended-exp}, we will show some experiments to illustrate the effect of this technique
when changing the selectivity. 

\subsubsection{Other Query Patterns}

We examine some other popular patterns against other systems that
support a high-level language. We see that LFTJ is fastest on cyclic
queries, while \ms is the fastest on acyclic queries. We then consider
queries that contain cyclic and acyclic components.

\begin{itemize}
\item\textbf{4-cycle} Table~\ref{fig:results-cyclic} shows the 4-cycle
  results. PostgreSQL and MonetDB perform are slower by orders of
  magnitude, similar to the results for \{3,4\}-clique. LFTJ is
  significantly faster than Minesweeper on this cyclic query.

\item\textbf{\{1,2\}-tree} LFTJ is the fastest for the 1-tree query, but
has trouble with the 10\% Orkut experiment. Minesweeper handles all
datasets without issues for 1-tree, and is faster than LFTJ, which
times out on many experiments. PostgreSQL and MonetDB both timeout on
almost all of the 2-tree experiments. With a few exceptions,
PostgreSQL does perform well on the 1-tree experiments. Minesweeper
benefits from instance optimality on this acyclic query.
\end{itemize}

We consider the $i$-lollipop query combines the $i$-path and $(i+1)$-clique
query for $i\in \{2,3\}$, so predictably PostgreSQL and MonetDB do very poorly. LFTJ does
better than Minesweeper, which suffers from the clique part of the
query, but LFTJ on the other hand suffers from the path aspect and
times out for most bigger data sets. The hybrid algorithm presented in 
Section~\ref{subsec:lollipop} outperforms both and the results are illustrated 
in Table~\ref{fig:results-acyclic}. This may be an interesting research direction.

\subsubsection{Extended Experiments}
\label{extended-exp}
\begin{figure}[H]
\includegraphics[width=3.3in]{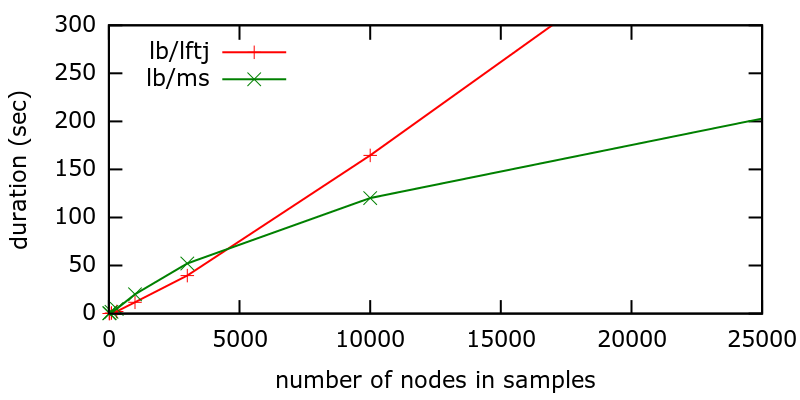}
\vskip-1em\caption{3-path on LiveJournal with samples of N nodes}
\label{fig:results-3path-livejournal-select-scaling}

\includegraphics[width=3.3in]{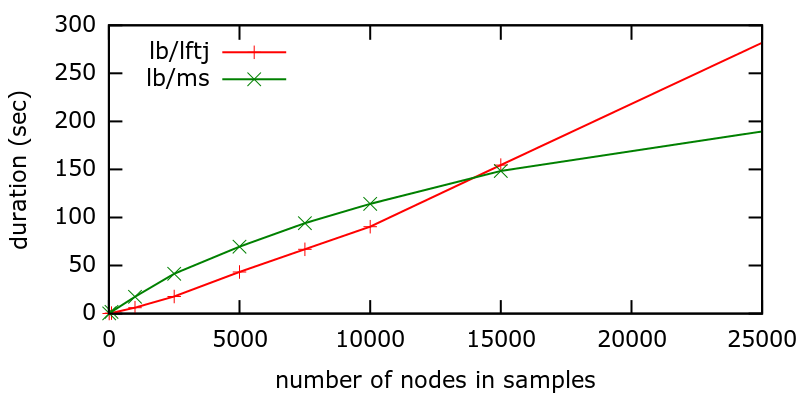}
\vskip-1em\caption{3-path on Pokec with samples of N nodes}
\label{fig:results-3path-pokec-select-scaling}

\includegraphics[width=3.3in]{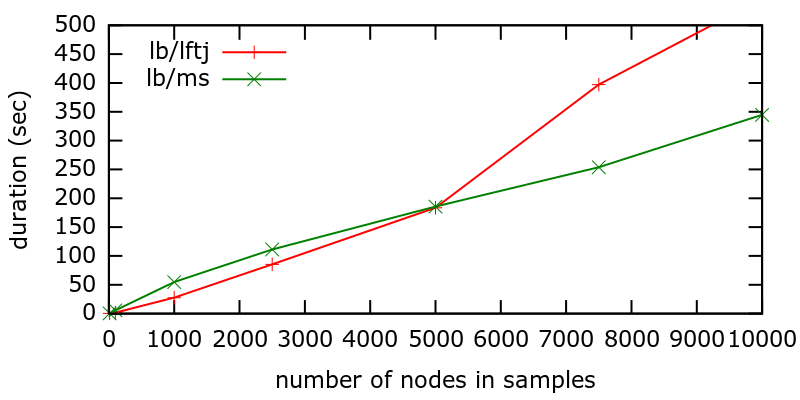}
\vskip-1em\caption{3-path on Orkut with samples of N nodes}
\label{fig:results-3path-orkut-select-scaling}
\end{figure}
To illustrate the fact that Minesweeper has caching mechanism to avoid 
doing redundant work, Figure~\ref{fig:results-3path-livejournal-select-scaling},
\ref{fig:results-3path-pokec-select-scaling}, and
\ref{fig:results-3path-orkut-select-scaling} compare the performance
of the algorithms when 3-path is executed with increasingly larger
node samples.
\begin{figure}[t]
\includegraphics[width=3.3in]{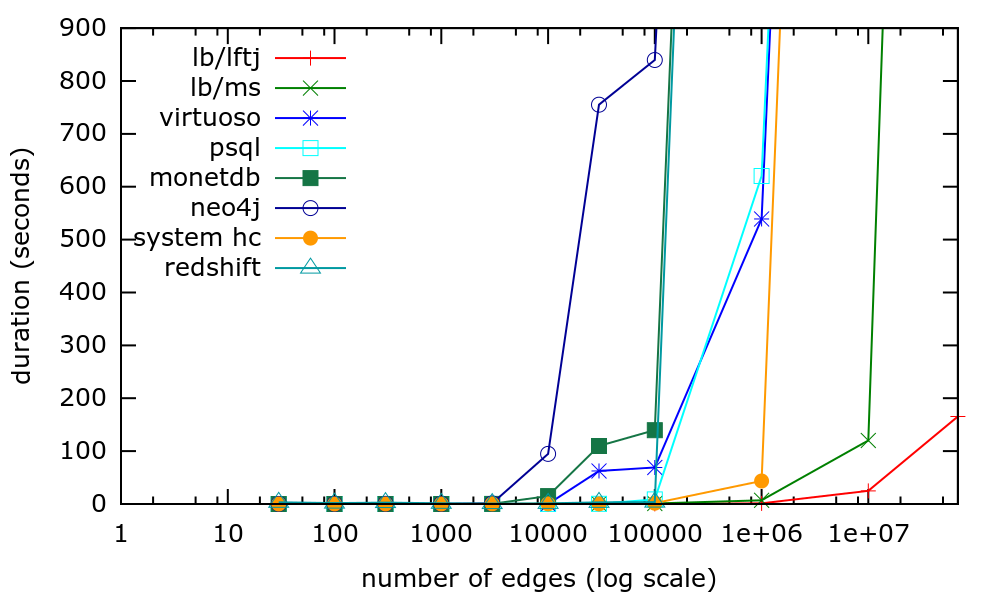}
\vskip-1em\caption{Duration of 3-clique on LiveJournal subset of N edges}
\label{fig:results-3clique-trend}  

\includegraphics[width=3.3in]{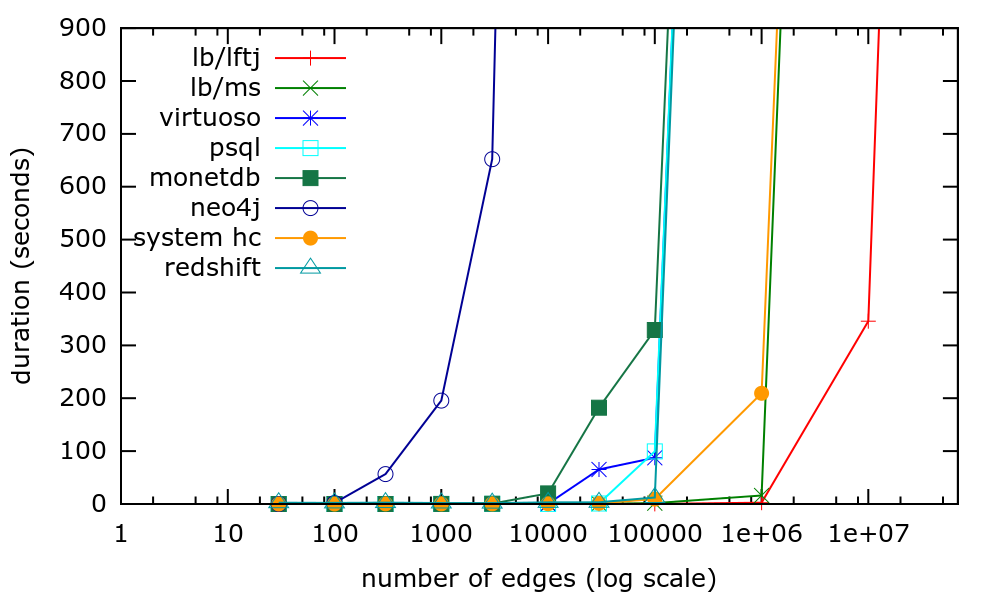}
\vskip-1em\caption{Duration of 4-clique on LiveJournal subset of N edges}
\label{fig:results-4clique-trend} 
\end{figure}

\paragraph*{Scaling Behavior} To understand the scaling issues better, we eliminate the variability
of the different datasets and execute a separate experiment where we
gradually increase the number of edges selected from the LiveJournal
dataset with a timeout of 15 minutes. This time we also include
RedShift and System HC. The results are shown in
Figures~\ref{fig:results-3clique-trend}
and~\ref{fig:results-4clique-trend}. This analysis shows that
conventional relational databases (and Neo4J) do not handle this type
of graph query even for very small data sets. Virtuoso is the best
after the optimal joins. Optimal joins can handle subsets of two
orders of magnitude bigger, and LFTJ supports an order of magnitude
bigger graphs than Minesweeper.

\paragraph*{Summary} LogicBlox using the LFTJ or
Minesweeper algorithms consistently out-performs other systems that
support high-level languages. LFTJ performs fastest on cyclic
queries and is competitive on acyclic queries (1-tree) or queries with
very high selectivity. Minesweeper works best for all other acyclic
queries and performs particularly well for queries with low
selectivity due to its caching.

\section{Conclusion}
\label{sec:conclusion}

Our results suggest that this new class of join algorithms allows a
fully featured, SQL relational database to compete with (and often
outperform) graph database engines for graph-pattern matching. One
direction for future work is to extend this benchmark to recursive
queries and more graph-style processing (e.g., BFS, shortest path,
page rank). A second is to use this benchmark to refine this still
nascent new join algorithms. In the full version of this paper, we
propose and experiment with a novel hybrid algorithm between LFTJ and
\ms. We suspect that there are many optimizations possible for these
new breed of algorithms.

\eject
{\small
\bibliographystyle{abbrv}
\bibliography{minesweeper}
}


\newpage
\onecolumn\appendix
\section{\agm bound}
\label{app:agm}

Given a join query $Q$ whose hypergraph is $\calH(Q) = (\calV, \calE)$,
we index the relations using edges from this hypergraph.
Hence, instead of writing $R(\vars(R))$, we can write $R_F$, 
for $F\in\calE$.

A {\em fractional edge cover} of a hypergraph $\calH$ is a point
$\mv x = (x_F)_{F\in\calE}$ in the following polyhedron: 
\[ \left\{ \mv x \suchthat \sum_{F: v \in F} x_F \geq 1, \forall v \in \calV,
\mv x \geq \mv 0 \right\}. \]

Atserias-Grohe-Marx \cite{AGM08} and Grohe-Marx \cite{GM06}
proved the following remarkable inequality, which shall be 
referred to as the {\em \agm's inequality}.
For any fractional edge cover $\mv x$ of the query's hypergraph, 
\begin{equation}
\label{eqn:AGM}
 |Q| = |\Join_{F\in\calE}R_F| \leq \prod_{F\in\calE} |R_F|^{x_F}. 
\end{equation}
Here, $|Q|$ is the number of tuples in the (output) relation $Q$.

The optimal edge cover for the \agm bound depends on the relation sizes.
To minimize the right hand side of \eqref{eqn:AGM}, we can solve the
following linear program:
\begin{eqnarray*}
\min && \sum_{F\in \calE} (\log_2 |R_F|) \cdot x_F\\
\text{s.t.} && \sum_{F: v \in F} x_F \geq 1, v \in \calV\\
&&\mv x \geq \mv 0
\end{eqnarray*}
Implicitly, the objective function above depends on the database instance
$\calD$ on which the query is applied. 
We will use $\agm(Q)$ to denote the best \agm-bound for the input instance
associated with $Q$.
\agm showed that the upper bound is essentially tight in the sense that there
is a family of database instances for which the output size is 
asymptotically the same as the upper bound. Hence, any algorithm whose runtime 
matches the \agm bound is optimal in the worst-case.
\section{Tuning parameters per system}
\label{tuning-param}
\begin{center}
{\sf\small
\def\arraystretch{1.2}
\begin{tabular}{l|l|l|r}
  \multicolumn{3}{l|}{tuning parameter}
  & value
  \\

\hline
\multicolumn{2}{l}{psql}     & temp\_buffers & 2GB   \\
\multicolumn{2}{l}{}         & work\_mem     & 256MB \\

\hline
\multicolumn{2}{l}{virtuoso} & NumberOfBuffers & 2380000 \\
\multicolumn{2}{l}{}         & MaxDirtyBuffers & 1750000 \\
\multicolumn{2}{l}{}         & TransactionAfterImageLimit & 99999999 \\

\hline
\multicolumn{2}{l}{neo4j}    & neostore.nodestore.db.mapped\_memory & 500M  \\
\multicolumn{2}{l}{}         & neostore.relationshipstore.db.mapped\_memory & 3G \\
\multicolumn{2}{l}{}         & neostore.propertystore.db.mapped\_memory & 500M \\
\multicolumn{2}{l}{}         & wrapper.java.initmemory (-Xms) & 16384 \\
\multicolumn{2}{l}{}         & wrapper.java.maxmemory (-Xmx) & 16384 \\
\multicolumn{2}{l}{}         & wrapper.java.additional.1 & -Xss1m  \\

\hline
\multicolumn{2}{l}{graphlab} & ncpus & 8 \\

\end{tabular}
}
\end{center}

\end{document}